\documentclass[prb,amsmath,amssymb]{revtex4-2} 

\usepackage{graphicx}% Include figure files
\usepackage{bm}% bold math
\newcommand{\ve}[1][K]{\mathbf{#1}}

\usepackage{amsmath}
\usepackage{amssymb}
\usepackage{latexsym}
\usepackage{multirow}
\usepackage{xcolor}
\usepackage{hyperref}
\hypersetup{colorlinks=true,linkcolor=blue,citecolor=blue}
 
 \newcommand{\dif}[1][d]{\mathrm{#1}}
 
\begin{document}

%\title{Theory of competitive first passage for non-Markovian Gaussian stochastic processes }
\title{Evidence and quantification of memory effects in competitive first passage events}

% MAYBE THE SHORT VERSION 
%\title{Memory effects in competitive first passage events}

% OR OTHER TITLE ! 

\author{M. Dolgushev$^{1}$, T. V. Mendes$^{2}$, B. Gorin$^{2}$, K. Xie$^{2}$, N. Levernier$^{3}$, O. B\'enichou$^1$,  H. Kellay$^{2}$,    R. Voituriez$^{1,4}$, T. Gu\'erin$^{2}$}
\affiliation{$^{1}$ Laboratoire de Physique Th\'eorique de la Mati\`ere Condens\'ee, CNRS/Sorbonne University, 
 4 Place Jussieu, 75005 Paris, France}
\affiliation{$^{2}$Laboratoire Ondes et Mati\`ere d'Aquitaine, CNRS/University of Bordeaux, F-33400 Talence, France}
\affiliation{$^{3}$ CINaM, CNRS / Aix Marseille Univ, Marseille, FRANCE}
 \affiliation{$^{4}$ Laboratoire Jean Perrin, CNRS/Sorbonne University, 
 4 Place Jussieu, 75005 Paris, France }

\bibliographystyle{naturemag}
\date{\today}

\begin{abstract} 
\textbf{Splitting probabilities quantify the likelihood of a given outcome out of competitive events. This key observable of random walk theory, historically introduced as the  gambler's ruin problem, is well understood for memoryless (Markovian) processes. However, in complex systems such as polymer fluids, the motion of a particle should typically be described as a process with memory, for which  splitting probabilities are much less characterized analytically. Here, we introduce an analytical approach that provides the splitting probabilities for one-dimensional isotropic non-Markovian Gaussian processes with stationary increments, in the case of two targets. This analysis  shows that splitting probabilities are controlled by the out of equilibrium trajectories observed after the first passage. This is directly evidenced in a prototypical experimental reaction scheme in viscoelastic fluids. These results are extended to $d$-dimensional processes in large confining volumes, opening a path towards the study of competitive events in complex media.}
\end{abstract}

\maketitle

\section*{Introduction}
\begin{figure}[t!]
 \centering
\includegraphics[width=12cm]{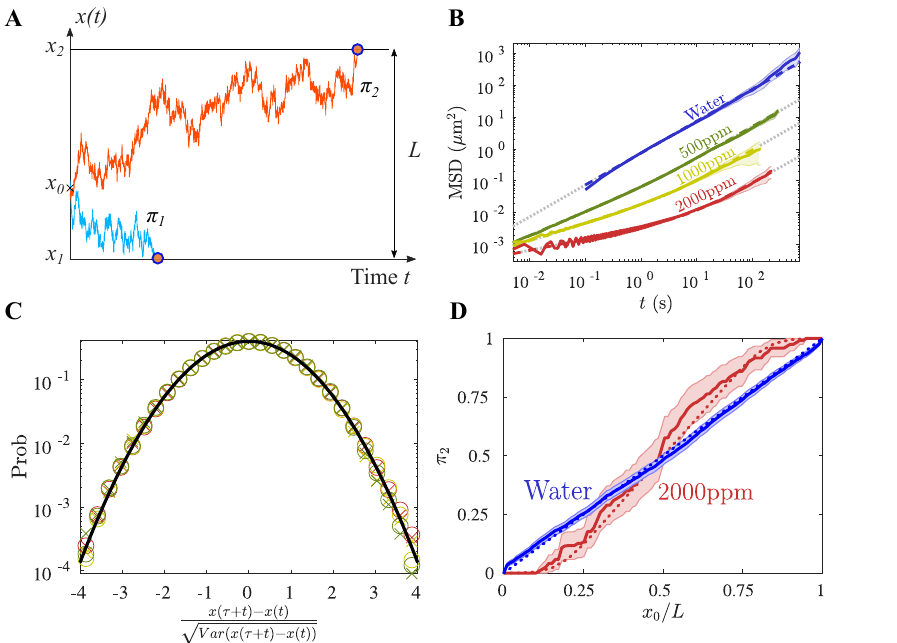}
\caption{{\textbf{Competitive events and non-Markovian motion.}}
\textbf{A.} Sketch of the problem of competitive events investigated in this paper: in presence of two targets, for a random walk, what is the probability to hit one target before the other? \textbf{B.} Experimental Mean Square Displacement (MSD) of tracer particles of $1\mu$m diameter in viscoelastic fluids, with polymer concentrations (from top to bottom) $c=0$ ppm (blue), $c=500$ ppm (green), $c=1000$  ppm (yellow) and $c=2000$ ppm (red). The dashed lines are fits using Eq.~(\ref{ShapeMSD}). \textbf{C.} normalized histograms of the increments $x(t+\tau)-x(t)$ for $\tau=\tau_0$ (crosses) and $\tau=\tau_0/2$ (circles) for the three polymer concentrations of $\textbf{B}$ (with the same color code).  Here $\tau_0$ is the memory time of the solution defined in Eq.~(\ref{ShapeMSD}), with $\tau_0\simeq 1.48$ s for $c=500$ppm, $\tau_0\simeq 2.9$ s for $c=1000$ ppm and $\tau_0\simeq7$ s for $c=2000$ ppm. The black line is the density of a normalized Gaussian, $p(x)=e^{-x^2/2}/\sqrt{2\pi}$. 
\textbf{D} Values of the splitting probabilities measured for beads in water (blue curves) and in viscoelastic fluids at $c=2000$ ppm (red) in our experiments and theory (dotted lines), for  $L=0.6\mu$m. {Error bars are twice the standard deviation of the mean, calculated with $n=116$ recorded first passage events. }
}
\label{Fig1}
\end{figure}

Which will you reach first: fortune or ruin? It is ubiquitous that the fate of a system depends on which of a finite set of possible outcomes is realized first, see Fig. \ref{Fig1}\textbf{A}. An historical example is provided by the ``gambler's ruin problem'', in which one wishes to know the risk for a gambler  to go bankrupt before making a given profit \cite{feller68}. A form of this problem can be traced back to Pascal, and quantifying the risk of ruin has become a classical problem in financial mathematics \cite{bouchaud2018trades}. In fact, competitive events appear in various domains and under different names; examples include the fixation probability of a mutant in the context of population dynamics~\cite{AlfredMoranBook1962}, or the nucleation probability in the classical nucleation theory of phase transitions \cite{richard18}. In polymer physics, this question emerges in the problems of DNA melting \cite{oshanin09,lubensky02}, protein and RNA hairpin folding \cite{best05,chodera11}, polymer translocation through a small pore \cite{lua05} and crystallization \cite{guttman82,mansfield88}, polymer adsorption/desorption kinetics \cite{wittmer94} and, related to it, cell adhesion \cite{jeppesen01,bell17}. 
{Estimating conditional splitting probabilities is also important to determine entropy production~\cite{van2022thermodynamic,hartich2023violation,blom2024milestoning,martinez2019inferring}}. Another important field of applications, to which we will refer through this paper, is given by competing (also known as parallel or concurrent) chemical \cite{espenson95}, biochemical \cite{hansen19} or photochemical \cite{motti19} reactions.

The key quantity characterizing generic competitive events is the \textit{splitting probability}, i.e. the probability, for a random process, of realizing first a given event before several others could occur, and as such belongs to the class of first-passage observables. Most of available theoretical methods to determine splitting probabilities are limited to 1--dimensional Markovian processes~\cite{VanKampen1992,Redner:2001a}. Recent advances have considered  the extension to higher dimensions for  Brownian random walks ~\cite{dobramysl2020triangulation,cheviakov2011optimizing,bressloff2020target,bressloff2021first,condamin2007random,condamin2006exact,Condamin2005} and  general Markovian processes (\textit{i.e.}, processes without memory)~\cite{condamin2008probing,benichou2014first}. 

However, memory effects are essential in complex systems since they emerge as soon as the evolution of the random walker, or of the reaction coordinate, arises from interactions with other (possibly hidden) degrees of freedom. For example, the motion of a monomer in a macromolecule \cite{Panja2010,bullerjahn2011monomer,sakaue2013memory}, or that of a particle in a crowded narrow channel \cite{wei2000single}, display strong memory effects. Another well known experimental example of a non-Markovian process (to be studied below) is the motion of a tracer bead in a viscoelastic solution \cite{mason1997particle,mason1995optical,squires2010fluid,furst2017microrheology}, for which examples of mean square displacement (MSD) functions are shown on Fig. \ref{Fig1}\textbf{B}, which clearly display several temporal regimes and strongly differ from Brownian motion, as expected in such complex fluids \cite{vanZanten2004brownian}. The Gaussian nature of this process, as seen on Fig.~\ref{Fig1}\textbf{C}, together with the temporal non-linearity of the MSD, ensures that the observed process is indeed non-Markovian \cite{kallenberg1997foundations}. 

However, a general theory to quantify the impact of memory effects on splitting probabilities is lacking. 
Indeed, in the context of first-passage problems for non-Markovian processes,  theoretical approaches have mainly been limited to the case of single targets \cite{guerin16,ReviewBray,levernier2019survival,dolgushev15,delorme2015maximum,sadhu2018generalized,Levernier2022Everlasting,walter2021first,levernier2020kinetics}. In the case of two targets, the prediction of splitting probabilities is limited to 1--dimensional processes, in a few specific examples \cite{masoliver1986first,bicout2000absorption} or for scale invariant processes using scaling arguments~\cite{majumdar10} or perturbative methods~\cite{wiese19}. 

Here, we introduce a general non-perturbative formalism to predict the outcome of competitive events for  the wide class of {non-smooth isotropic} Gaussian processes {with stationary increments (see below for definition), in the case of two targets}. 
Strikingly, on the basis of a prototypical experimental reaction scheme with tracer beads in viscoelastic fluids, 
we provide a direct experimental  evidence of the impact of memory effects on competitive reactions [see Fig.~\ref{Fig1}\textbf{D} for illustration], in agreement with our theoretical predictions, while so far experimental observations of first passage properties of non-Markovian processes have been limited to persistence exponents \cite{ReviewBray,wong2001measurement} or passage over barriers \cite{ginot2022barrier,ferrer2021fluid}.    In particular, our observations   provide a direct experimental proof that the state of the system (constituted by the random walker and the additional degrees of freedom of its environment) at the first passage event is not an equilibrium state. Our theory extends  to  dimensions higher than 1, providing a path towards the understanding of competitive diffusion limited reactions in complex systems.

\section*{Results}

\textit{General formalism.} We first consider a random walker of position $x(t)$ evolving in continuous time $t$ in a one-dimensional space (the generalization to higher-dimensions  will be considered afterwards). We assume that the random walk is symmetric (no privileged direction) and that the increments are stationary (no aging). The initial position is $x_0$. We also assume that the process $x(t)$ is continuous (no jumps) and non-smooth (with formally infinite velocity, as in the case of overdamped processes and in particular of Brownian motion), and that it has Gaussian statistics. With these hypotheses, the process is fully characterized by its average ($\langle x(t)\rangle=x_0$) and the MSD function $\psi(t)=\langle [x(t)-x_0]^2\rangle$. Last, we assume that, at long times only, $\psi(t)$ behaves as $\psi(t)\simeq \kappa \ t^{2H}$ with $\kappa>0$ and $0<H<1$. {The hypothesis $H>0$ ensures that the particle does not remain trapped around a given position, and  the condition $H<1$ ensures that the correlation function of the increments decays at long times.} The process is therefore assumed to be diffusive ($H=1/2$), subdiffusive ($H<1/2$) or superdiffusive $(H>1/2)$ at long times. With these hypotheses one describes a large class of non-Markovian random walks, and in particular diffusion of beads attached to macromolecules~\cite{Panja2010,Panja2010a,bullerjahn2011monomer,blumen2004generalized,Dolgushev2009}, or moving in viscoelastic fluids~\cite{mason1997particle,mason1995optical} or crowded narrow channels~\cite{wei2000single}, etc. {We stress that the above assumptions   refer to the dynamics  of the random walker in the absence of targets.}

We now consider   two perfectly absorbing targets at positions $x_1=0$ and $x_2=L$ (with $0<x_0<L$). The random walk ends whenever one of these two regions is reached and we aim to calculate $\pi_i$, the probability that the target $i\in\{1;2\}$ is reached first. 
In the single target problem \cite{guerin16,levernier2019survival,levernier2020kinetics}, it was previously shown that a key quantity to predict first passage statistics is the average trajectory after first contact, if the random walker were allowed to continue its motion. We thus introduce $\mu_1(t)$ and $\mu_2(t)$, the average trajectories at a time $t$ after a first contact with targets $1$ and $2$, respectively. The following probabilistic argument enables one to understand why $\mu_1$ and $\mu_2$ are inherently linked to the splitting probabilities. At long times, the average of $x(t)$ (without targets) is clearly $x_0$, but on the other hand the average of $x$ can be computed by partitioning over the first contacts with each of the targets, leading to 
 \begin{align}
 x_0=\lim_{t\to\infty} [\pi_1 \mu_1(t)+\pi_2 \mu_2(t)]. \label{SmartArg}
 \end{align}
Note that, for the (Markovian) Brownian motion, $\mu_1=0$ and $\mu_2=L$ so that the above argument leads to the well-known result $\pi_2=x_0/L$. 
In the more general case of non-Markovian processes, this equations is key to evaluate $\pi_1$ (and $\pi_2=1-\pi_1$), but requires the knowledge of $\mu_1$ and $\mu_2$. A self-consistent equation for $\mu_1$ and $\mu_2$ can be obtained by assuming that the statistics of trajectories after a first contact is Gaussian, with the same covariance as that of the original process (see Supplementary Materials (SM), Section I). These assumptions are well supported by simulations (Fig.     \ref{fig:checkHist} in SM, Section II). The equations for $\mu_1,\mu_2$ are, for $i\in\{1,2\}$:
\begin{align}
0&=\sum_{j=1}^2 \pi_j \int_0^\infty dt 
 \Big(p(x_i,t) (x_0-x_i) [1-M(t,\tau)] \nonumber\\
&-  q_j(x_i,t) \{\mu_j(t+\tau)-x_i - [\mu_j(t)-x_i]M(t,\tau)  \}   \Big), \label{SelfCons}
\end{align}
with
\begin{align}
&M(t,\tau)=[\psi(t+\tau)+\psi(t)-\psi(\tau)]/[2\psi(t)], \nonumber\\
&q_j(x_i,t)=\frac{e^{-\frac{(x_i-\mu_j(t))^2}{2\psi(t)}}}{[2\pi\psi(t)]^{1/2}},\  \ p(x_i,t)=\frac{e^{-\frac{(x_i-x_0)^2}{2\psi(t)}}}{[2\pi\psi(t)]^{1/2}}. \label{Propag}
\end{align}
{Here, $p(x_i,t)$ is the value of the probability density function (PDF) of the initial process at the position $x_i$, while $q_j(x_i,t)$ is the PDF of observing a particle at position $x_i$ at a time $t$ after the first contact, given that the target $j$ is hit first ; {note that these quantities are defined for the process in infinite space.} Eq.~(\ref{SelfCons}) expresses the fact that the conditional average of $x(t+\tau)$, given that $x(t)=x_i$, averaged over $t$, can also be calculated by partitioning over events where the target $j$ is hit first in terms of averaged trajectories after the first contact; the function $M$ comes from general expression of conditional averages for Gaussian processes \cite{Eaton1983}.}
The above equations  (\ref{SelfCons}) can be solved numerically to evaluate $\mu_1(t)$ and $\mu_2(t)$, and therefore $\pi_1$ and $\pi_2$ by using Eq.~(\ref{SmartArg}), for arbitrary $\psi(t)$.  

\begin{figure}
 \centering
\includegraphics[width=14cm]{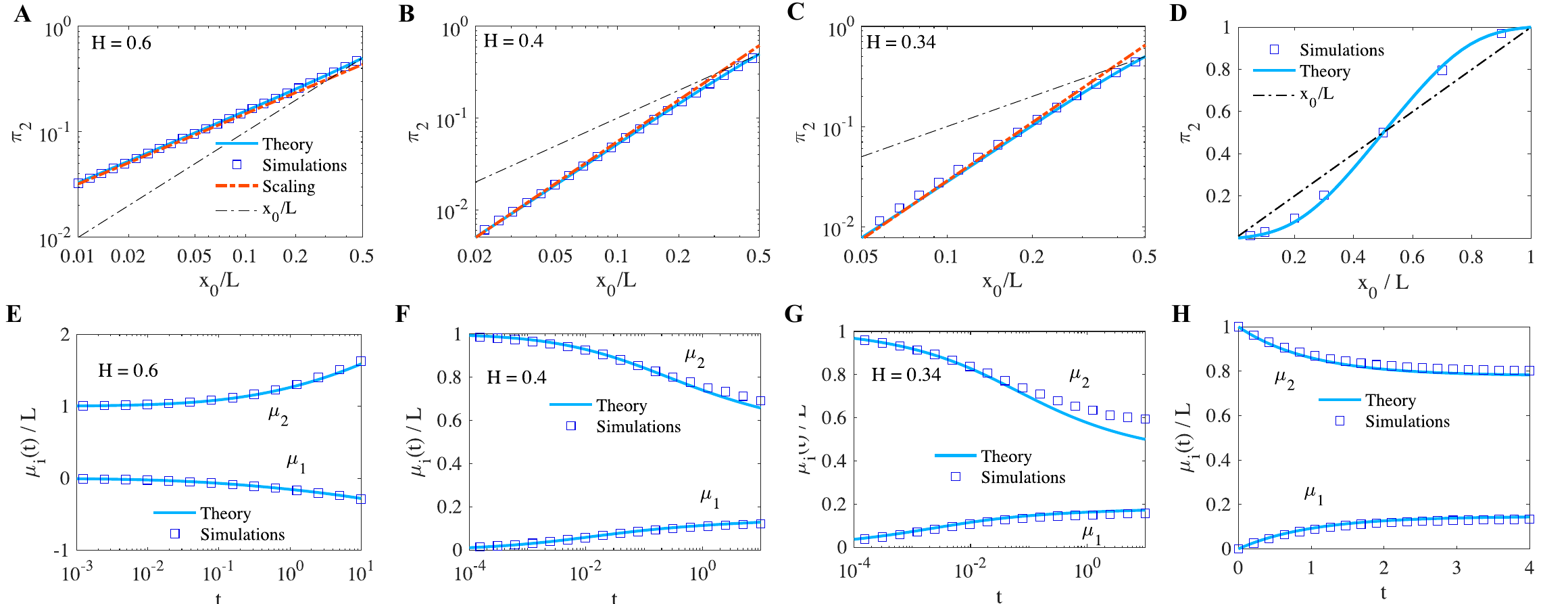}
\caption{{\textbf{Splitting probabilities and mean trajectories after the first passage for one dimensional processes}}. \textbf{A}: Splitting probability for a superdiffusive fBM with $H=0.6$. Symbols are simulation results  obtained with the circulant matrix algorithm \cite{Dieker2004,dietrich1997fast} (statistical error is smaller than symbol sizes). The blue continuous line is the theoretical prediction, obtained by numerically solving  Eqs.~(\ref{SmartArg}) and (\ref{SelfCons}). The red dashed line is the scaling (\ref{ScalingSmallx0}) with the prefactor $A_H$ predicted by our theory. The black dashed line is the formula $\pi_2=x_0/L$, obtained by setting $\mu_1=0$ and  $\mu_2=L$, that overlooks non-Markovian effects. 
\textbf{B},\textbf{C}: Splitting probability for  sub-diffusive fBMs ($H=0.4$ and $H=0.34$), with the same color code as in \textbf{A}. 
\textbf{D}: Splitting probabilities for  a non-scale invariant process with MSD {$\psi(t)=t+B(1-e^{-t})$ with $B=10$} (bidiffusive process), when the separation between the targets is $L=20$ (in dimensionless units), with the same color code as in \textbf{A-C}. 
\textbf{E,F,G,H}: Average trajectories $\mu_i(t)$ in the future of first passage events as measured in simulations (symbols) and predicted by our theory (lines), for $x_0/L=$ \textcolor{black}{0.208}, for the processes corresponding to \textbf{A}-\textbf{D}, respectively. In \textbf{E}-\textbf{G}, the time $t$ is in units of $(L/\sqrt{\kappa})^{1/H}$. {In \textbf{A-C} only the range $x_0<L/2$ is shown, the values of $x_0>L/2$ can be deduced by symmetry. The same figure, with linear scales, is shown in Fig.~S2.}
}
\label{FIG_FBM}
\end{figure}

Several comments are in order.
First, for weakly non-Markovian processes, i.e. when one can write $\psi(t)=2D t+\varepsilon \psi_1(t)$, with $\varepsilon $ a small parameter, our theory provides \textit{exact results at order $\varepsilon$} (SM, Section V), and we obtain explicit formulas for the splitting probability at this order for any $\psi_1$, which agree with  the result of Ref.~\cite{wiese19} found with other methods in the particular case where $\psi(t)= \kappa \ t^{2H}$ with $H\to1/2$. Second, we compare the predictions of our formalism to simulation results for several  stochastic processes, including strongly non-Markovian processes. To this end we consider two paradigmatic examples of Gaussian non-Markovian processes: (1) the fractional Brownian motion (fBM) with MSD $\psi(t)=\kappa\ t^{2H}$, which is scale invariant (at all times), this model displays  long range  memory effects  and appears in various fields; in particular it can describe the dynamics  of a monomer of an infinite polymer chain  \cite{Panja2010,bullerjahn2011monomer,sakaue2013memory}, or of a tagged particle in single file diffusion  \cite{wei2000single}; (2) the ``bi-diffusive'' process, with MSD $\psi(t)=t + B (1-e^{- t})$ (in dimensionless variables, {with $B>0$}). This process appears when $x(t)$ is driven by the sum of a white noise and a colored one, with only one relaxation time, as in a Maxwell fluid \cite{grimm2011brownian}. 
Numerical simulations of these processes show a quantitative agreement with our theoretical predictions for $\pi_2$  in all cases (see Fig.~\ref{FIG_FBM}). \textcolor{black}{Note that, in our calculations, we used} $\pi_2=[(x_0-\mu_1)/(\mu_2-\mu_1)]_{t\to\infty}$ \textcolor{black}{and we checked that the convergence to the limit is fast enough to enable the efficient  numerical determination of} $\pi_2$ (even when $\mu_1$ and $\mu_2$ both diverge\textcolor{black}{, see Fig.~  \ref{figFBMLinear} in SM}). 
Of note,  the Markovian prediction $\pi_2=x_0/L$ can either strongly underestimate [Fig.~\ref{FIG_FBM}\textbf{A}] or overestimate  [Fig.~\ref{FIG_FBM}\textbf{B},\textbf{C}]  $\pi_2$, with  an obviously incorrect scaling behavior at small $x_0$, while our approach remains quantitative in this regime.

Third, we analytically examine  the case of scale invariant processes, $\psi(t)=\kappa\ t^{2H}$, for which the dependence on the geometric parameters $x_0$ and $L$ can be extracted. For $x_0\ll L$, we find
\begin{align}
\pi_2 \underset{{x_0\ll L}}{\simeq} A_H (x_0/L)^{1/H-1},\label{ScalingSmallx0}
\end{align}
where the prefactor $A_H$ is determined below. Of note, this scaling behavior is consistent with that obtained from scaling arguments  in Ref.~\cite{majumdar10} for 1-dimensional processes, and extended in Ref.~\cite{levernier2018universal} to higher dimensions, in agreement with earlier predictions for Markovian scale invariant processes~\cite{condamin2008probing}.
Our approach in addition provides the quantitative determination of  the prefactor  $A_H$, unknown so far. Indeed, when $x_0/L\ll 1$, $\mu_2$ is not expected to depend on $x_0$, whereas $\mu_1$ varies at two time scales: the typical time to travel a distance $x_0$, equal to $\tau_1=(x_0/\sqrt{\kappa})^{1/H}$, and the time to travel a distance $L$, equal to $\tau_2=(L/\sqrt{\kappa})^{1/H}$. We find that the structure of $\mu_1$ in terms of matched asymptotic expansions is 
\begin{align}
&\mu_1(t,x_0)\simeq 
\begin{cases} 
x_0 \ \mu_\infty(t/\tau_1) & (t\ll \tau_2)\\
x_0 - x_0 \left(\frac{x_0}{L}\right)^{\frac{1}{H}-2} \chi(t/\tau_2) &(t\gg \tau_1)
\end{cases}\\
&\mu_2(t,x_0)\simeq L\ m_2(t/\tau_2)
\end{align}  
where $\mu_\infty,m_2$ and $\chi$ are dimensionless scaling functions satisfying {a set of } equations identified in SM [Section III, {Eqs.~(S25), (S26) and (S29)}]. Using Eq.~(\ref{SmartArg}), it is clear that our formalism yields the scaling (\ref{ScalingSmallx0}) and provides the value of the prefactor  $A_H=(\chi/m_2)_{t\to\infty}$, in excellent agreement with simulation results (see Fig.~\ref{FIG_FBM}) { in the regime $x_0\ll L$}.

\textit{Experiments. } We have experimentally measured first passage events for non-Markovian processes by observing the motion of micrometer sized beads in viscoelastic large polymer weight solutions (details on experiments can be found in Methods and SM, Section IV). {The beads are far from the confining boundaries containing the polymer fluid, and we focus on their motions along the  $x$-axis, which were} tracked by using  optical microscopy. This type of experimental set-up is standard in microrheology \cite{waigh2005microrheology,mason1997particle,mason1995optical,squires2010fluid,furst2017microrheology} but is usually used to measure viscoelastic parameters and not first passage properties. The motion of the beads can be interpreted as obeying an overdamped Generalized Langevin Equation (GLE)
\begin{align}
\int_0^t dt' K(t-t')\dot{x}(t')=\xi(t), \label{GLE}
\end{align}
where the Gaussian noise $\xi(t)$ has vanishing average and satisfies $\langle \xi(t)\xi(t')\rangle=k_BT K(\vert t-t'\vert )$. The measured MSDs in Fig.~1\textbf{C} typically display two regimes: one long time diffusive regime and one short time regime where one observes apparent anomalous diffusion. This suggests that, to account  for the observed trajectories, one can use a friction kernel of the form of
\begin{align}
K(t) = \frac{\gamma_0 \tau_0^{\alpha-1} e^{-t/\tau_0}}{\Gamma(1-\alpha) t^{\alpha}},
\end{align}  
where $\tau_0$ is the relaxation time of the polymer solution, $\gamma_0=\int_0^\infty dt K(t)$ is the long-time friction coefficient and $\alpha$ is the subdiffusion exponent at small times. For this memory friction kernel, the MSD reads 
\begin{align}\label{ShapeMSD}
&\psi(t)=\frac{2k_BT \tau_0 }{\gamma_0}f(t/\tau_0),\\ 
&f(y)= \left[ (y-\alpha+1)\gamma(\alpha,y)+y^{\alpha}e^{-y}\right]/\Gamma(\alpha), 
\end{align}
where $\gamma(\alpha,y)=\int_0^y t^{\alpha-1}e^{-t}dt$ is the lower incomplete gamma function. With this choice of kernel, the MSD displays a short time anomalous diffusive regime $\psi(t)\propto t^{\alpha}$ and a long time diffusive one, and $\tau_0$ is the crossover time between these regimes. The fits of experimental MSD curves show a good agreement,  as seen on Fig.~\ref{Fig1}\textbf{B}, and support this choice of function for $K$. We  checked  that the stochastic process $x(t)$ in the experiment is Gaussian [Fig.~\ref{Fig1}\textbf{C}] and unbiased  (SM, Section IV), as it should be if it is a realization of the GLE  (\ref{GLE}), and as implicitly assumed in all microrheology experiments. 

\begin{figure}[t!]
  \centering
\includegraphics[width=0.95\linewidth]{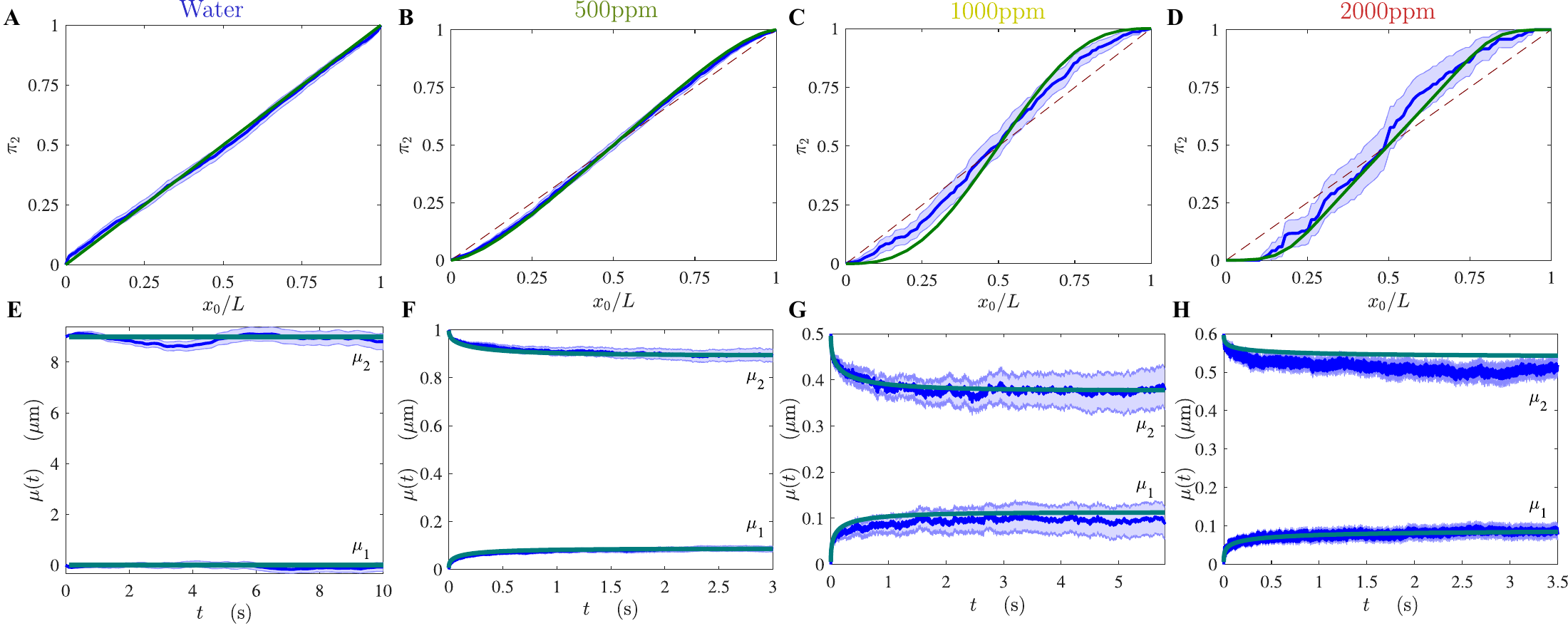} 
\caption{{\textbf{Experimental and theoretical splitting probabilities and mean trajectories after the first passage for a random walker in a viscoelastic fluid}.} 
\textbf{A-D}: Values of the splitting probabilities measured in experiments (blue lines, surrounded by estimates of  $95\%$ errorbars), compared with the prediction of our theory (green line). The polymer concentration is indicated on each graph. We also show  the result for Markovian diffusion $\pi_2=x_0/L$ (red dashed line). \textbf{E-F}:  Average trajectories $\mu_1$ and $\mu_2$ after the first passage to targets $1$ and $2$, as measured in the experiments (blue lines) and predicted by the theory (green lines). Parameters: 
\textbf{A},\textbf{E}: $c=0$ppm (water) and $L=9\mu$m, 
\textbf{B}, \textbf{F}: $c=500$ ppm and $L=1\mu$m, 
\textbf{C},\textbf{G}: $c=1000$ ppm and $L=0.5\mu$m, 
\textbf{D},\textbf{H}: $c=2000$ ppm and $L=0.6\mu$m. In \textbf{E}, $x_0=2\mu$m,  in \textbf{F-H},  $x_0=0.2\mu$m. 
{Error bars indicate $95\%$ confidence intervals and are calculated with the following number of recorded first passage events: $n=1001$ for $c=0$ ppm, $n=2771$ for $c=500$ ppm, $n=350$ for $c=1000$ ppm, and $n=116$ for $c=2000$ ppm.}}
\label{FigExp}
\end{figure}

Next, we investigated the first passage properties of a trajectory $x(t)$ starting at $x(t_0)=x_0$: we measured the variable $\eta$ defined as $\eta=1$ if a fictitious target at position $L$ is hit before the position $0$. To obtain enough statistics, we considered that $x(t_0)$ and $x(t_1)$ could be used as independent starting positions if the time elapsed between $t_1$ and $t_2$ is larger than $2\tau_0$ (or $2$s in water). Then, $\pi_2(x_0,L)$ was calculated as an ensemble average $\langle \eta\rangle$ over different starting positions on each trajectory for various bead trajectories. In our experiments, we made sure that the frame rate is large enough to consider that the typical distances traveled during each time step $dt$ is much smaller than $L$. The results for $\pi_2$ are displayed on Fig.~\ref{FigExp}(b-d). It is clear that for strongly viscoelastic solutions, for which $\alpha$ is the lowest, the curve $\pi_2(x_0)$ is very different from a straight line (which is the result for Markovian diffusion, see Fig.~\ref{FigExp}\textbf{A}), indicating that the probability to hit the closest target is increased by memory effects. This comes from the fact that, in polymer fluids, the motion of a tracer bead induces a delayed response of the surrounding polymer network, tending to bring it back to previously occupied positions, inducing a ``denser'' exploration of space and a larger probability to hit the closest targets. Furthermore, {for all parameters}, the experimental values of $\pi_2$ are in good agreement with our theoretical predictions {obtained by solving Eqs.~(\ref{SmartArg},\ref{SelfCons}) by using the previously fitted MSD (\ref{ShapeMSD})}.

In our experiments, we also measured the trajectories $\mu_1(t),\mu_2(t)$ followed by $x(t)$ after the first passage events, which are the hallmarks of non-Markovian effects in the theoretical approach above. 
These trajectories are displayed on Fig.~\ref{FigExp}(f-h) where it is clearly seen that, on average, $x(t)$ does not stay at $x=0$ or $x=L$ after the targets have been reached. 
These trajectories $\mu_1$ and $\mu_2$ are in quantitative agreement with their predicted values. In our experiments, the motion of a bead with  an equilibrium  initial condition would be  unbiased. In turn, our observation   that $\mu_i\neq x_i$ indicates that the state of the polymer fluid upon a first-passage event at a target is not an equilibrium state. Physically, this comes from the fact that the fluid  exerts a delayed response force tending to bring the bead back to its previously occupied positions, which, in our situation, are inside the interval $[0,L]$. This effect is, as expected, not present in solutions without polymers, see Fig.~\ref{FigExp}\textbf{E}. Our observations thus constitute a direct  experimental proof showing unambiguously that the state of a system (constituted by the bead and the surrounding polymer fluid) upon a first-passage event is not an equilibrium one, which is crucial to understand first passage properties.

\textit{Extension to higher spatial dimensions ($d>1$).} Importantly, our theory can be generalized  to higher dimensions, which is relevant to describe general competitive reactions. 
We denote the $d$--dimensional trajectory of the random walker by  $\ve[r](t)=(x_1(t),x_2(t),...x_d(t))$. We assume the presence of two targets of finite radius $a$ around locations $\ve[r]_1$ and $\ve[r]_2$, while $\ve[r]_0$ is the starting position. {The dynamics is assumed to take place in a confining volume, so that the PDF of positions in confinement $p^c(\ve[r],t)$ reaches a stationary value  $p_s^c(\ve[r])$ at large time ({Note that for $d=1$, the presence of confining walls, if beyond the targets, becomes irrelevant)}. We assume that, in the limit of large volume,  far from the boundaries, all $x_i(t)$} satisfy the hypotheses used for the motion of $x(t)$ in 1D. We also assume isotropy, so that the coordinates $x_i(t)$ are independent. 
We show in   SM (Section VI) that, {in the large volume limit},
\begin{align}
&\pi_1\textcolor{black}{=1-\pi_2}\textcolor{black}{\underset{V\to\infty}{\simeq}}\ \frac{h_{22}-h_{12}}{h_{22}+h_{11}-h_{21}-h_{12}} , \label{pi_all}
\end{align}
where 
\begin{equation}\label{h_ij}
h_{ij}=\int_0^{\infty}d t\ [q_j(\ve[r]_i,t)-p(\ve[r]_i,t)], 
\end{equation}
where $q_j(\ve[r],t)$ is the probability density function of the position $\ve[r]$ at a time $t$ after the first passage to target $j$, and $p$ is the probability density of the initial process {in unconfined space}. Note that, in our formalism, {in the large volume limit, }the propagators appearing in Eq.~(\ref{h_ij}) are evaluated by considering the dynamics in infinite space (without confining boundaries nor targets). The above equations generalize similar equations for Markovian processes~\cite{condamin2008probing,benichou2014first} to non-Markovian ones. Although the expressions are similar, the main difference from Markovian processes is that here the propagators have to be evaluated in the future of first passage events, which can strongly differ from the dynamics in the future of a stationary state. In $d=1$, Eq.~(\ref{pi_all}) is an alternative formula to estimate $\pi_i$ which however gives    results that are indistinguishable from those obtained with Eq.~(\ref{SmartArg}), see SM (Section VI, Fig.~\ref{FigSplittingWithhij}). 

\begin{figure}[h!]
 \centering
\includegraphics[width=15cm]{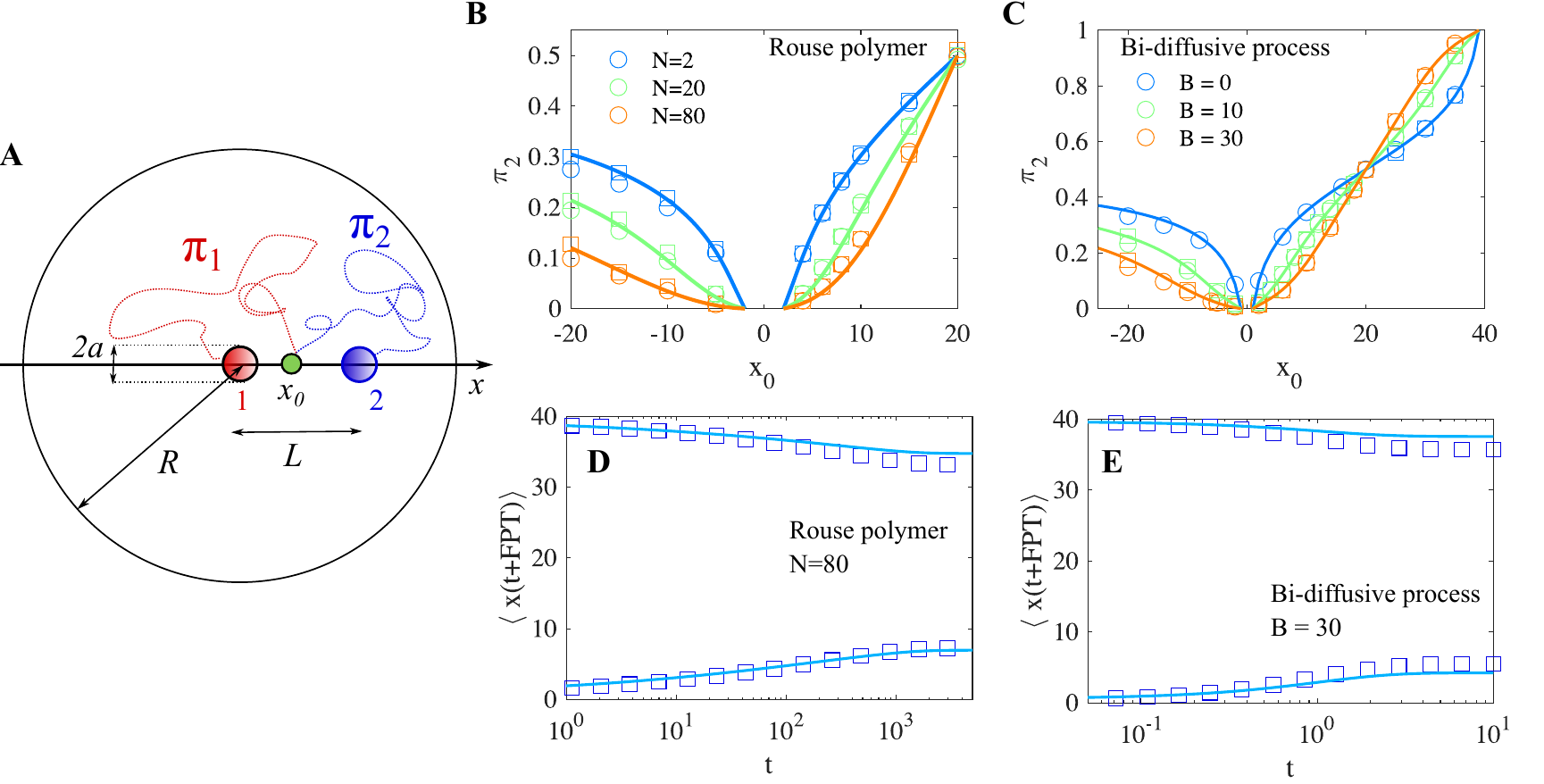} 
\caption{\textbf{Splitting probabilities in two dimensions.}
\textbf{A} Sketch of the competitive event problem in two dimensions, {when the confining volume is a disk of radius $R$ centered about the first target, with $L$ the distance between the targets and $a$ their radii. Here, $x_0$ is located on the line joining the centers of the targets and the position of the first target is $x_1=0$. 
\textbf{B} Splitting probability when the random walker is the first monomer of bead-spring (Rouse) polymer chain of $N$ monomers, whose dynamics obeys $ \partial_t\ve[x]_i=\ve[x]_{i+1}-2\ve[x]_i+\ve[x]_{i-1}+\ve[f]_i(t)$, with the prescription $\ve[x]_0=\ve[x]_1$ and $\ve[x]_N=\ve[x]_{N+1}$ at the ends, with $\langle f_i^\alpha(t) f_j^\beta(t')\rangle=2\delta_{\alpha\beta}\delta_{ij}\delta(t-t')$, with $i,j$ the monomers' indexes and $\alpha,\beta$ the spatial coordinates. Symbols are simulation results for $a=2$, $L=40$, and $N=2$ (blue), $N=20$ (green) and $N=80$ (red), circles are results for $R=80$ and squares for $R=160$. Lines represent  theoretical predictions [Eqs.~(\ref{pi_all}) and (\ref{h_ij})]. %Note that increasing memory effects (by increasing $N$) decreases the probability to reach the farthest target. 
\textbf{C} Splitting probability for the bidiffusive process with MSD $\psi(t)=t+B(1-e^{-t})$ for each coordinate. Symbols are simulation results for $a=1$, $L=40$, and $B=0$ (blue), $B=10$ (green) and $B=30$ (red), for  $R=90$ (circles) and $R=200$ (squares). 
\textbf{D,E} Projection of the mean trajectory after the FPT on the $x$-axis, when the target $1$ is hit before (lower curves) or after (upper curves) the target 2 for  \textbf{D} Rouse dynamics (with $N=80$, $x_0=15$, $R=160$, $a=2$, $L=40$) and  \textbf{E} for the bidiffusive process (with $B=30$, $x_0=14$, $R=90$, $a=1$, $L=40$). Lines are theoretical predictions. 
}
}
\label{Fig2Db}
\end{figure}

In $d-$dimensions, geometric effects are more difficult to take into account than in the 1D case, since it is necessary to evaluate in which direction the random walker moves after hitting a target. We address this problem by using the following approximations: 
(i) we work within a decoupling approximation, so that the statistics of paths in the future of the first passage to target $i$ is assumed to be the same as in the single target problem, 
(ii) we assume that the paths in the future of a first contact with target $i$ at hitting angle $\theta$ follow a Gaussian distribution, with mean   $\mu_i(t)$ oriented along this angle $\theta$, which it-self has  a distribution  $\Pi(\theta)\propto e^{\alpha \cos(\theta-\theta_0)}$. {This form for $\Pi(\theta)$ is a minimal ansatz of positive $2\pi$ periodic function, which is in good agreement with simulations (SM, Fig.~S6), and } we have written self-consistent equations for $\mu_i(t)$ and $\alpha$ (see SM, Section VI), which provide $\pi_2$. {Although the theory in $d$-dimensions involves more approximations than in the one-dimensional case, the} results shown in Fig.~\ref{Fig2Db} demonstrate that our theory  captures {well} the effects of memory on {the splitting probabilities for competitive reactions} in dimension higher than one. 

\section*{Discussion} 

Here we have presented a general theory that predicts the effect of memory on the outcome of competitive events, quantified by the splitting probability to reach one target before the other for {non-smooth isotropic} Gaussian stochastic processes {with stationary increments}. Our theory is exact at first non-trivial order for weakly non-Markovian processes and, beyond this perturbative regime, in quantitative agreement with both numerical simulations and experiments where the realization of the random walk is the motion of a tracer bead in a viscoelastic fluid.  Interestingly,  for this class of processes, the effect of memory is to increase the probability to hit the closest target (with respect to the Markovian prediction). This effect is also clear by looking at the case of Gaussian subdiffusive processes. This effect is strongly different from the case where subdiffusion arises from random jumps with heavy-tailed distributed waiting times, since the distribution of waiting times does not influence splitting probabilities \cite{condamin2008probing}. 
Our experiments also unambiguously demonstrate   that the state of the system (formed by the bead and the surrounding bath) at the first passage is not an equilibrium one {being conditioned to the random walker $x$ being at one of the targets},  as seen from the biased dynamics after the first passage events {(while the initial process is not biased); this aspect is actually intrinsically linked to splitting probabilities [see Eq.~(1)]. }  
Our theory can be extended to cover the case of reactions in spatial dimension higher than one {in the presence of a large confining volume,} opening a path to the study of the impact of memory effects on competitive reactions in complex media.

\section*{Materials and Methods}

\textbf{Preparation of polymer solutions, particles suspension and tracking methods - } Polyacrylamide (PAM, molecular weight Mw = 18 $\times$ 10$^6$ g/mol, from Polysciences) was dissolved in millipore water (18.2 M$\Omega$cm). 10 mM sodium chloride (NaCl, from Sigma-Aldrich) was added in the solution. The solution was then placed on a digital roller shaker (IKA Roller 6) for around 80 hours at a speed of 20-30 r.p.m. at room temperature to dissolve the polymer completely. The solution was stored in a fridge at 4-6 $^{\circ}$C. To minimize the effect of solution ageing, all solutions were used in this study within one month; we checked using rheological measurements that the solutions remained intact for such a period of time. Polystyrene particles from Invitrogen with diameter 1 $\mu$m were used in the experiments. 
Typically, 1.0 $\mu$L of original particle solution was added into a 2.5-3.0 mL polymer solution. The particles suspension was then mixed by using the digital roller shaker for 10 hours at 20-30 r.p.m. Experiments were carried out under a  darkfield inverted microscope (Zeiss AXIO Observer). The diluted particle suspension was sealed inside an adhesive incubation chamber (from Bio-Rad, 9 mm $\times$ 9 mm, 25 $\mu$L). The chamber was covered by a thin cover slide due to the limitation of the working distance of the darkfield condenser. The darkfield condenser was immersed in an optical oil over the microscope, and an objective (Olympus, SLMPLNx100) with magnification 100$\times$ and numerical aperture $0.6$ was used to visualise the particle motion. A motorized translation stage was used to capture microparticles from different areas. All videos were recorded by using a high resolution camera (Hamamatsu, OrcaFlash 4.0 C11440). To minimise the memory effect of the polymer solution, different places were chosen to record the particle motion. All particles were tracked with TrackPy (based on Python) \cite{crocker1996methods,Allan2021}. Additional details for the choice of experimental parameters, the estimator of the MSD and its variance, the check of the absence of global drift can be found in SM, Section IV. 
  
  \vspace{2cm}

\textbf{Acknowledgments:}  Computer time for this study was provided by the computing facilities MCIA (Mesocentre de Calcul Intensif Aquitain) of the Universit\'e de Bordeaux and of the  Universit\'e de Pau et des Pays de l’Adour. \\

\textbf{Funding:} T. G. and T-V. M. acknowledge the support of the grant ComplexEncounters, ANR-21-CE30-0020. 
 B. G, K. X. and H. K. thank the Institut Universitaire de France, LabEx AMADEUS ANR-10-LABEX0042-AMADEUS Université de Bordeaux and the ANR through project Lift,  ANR-22-CE30-0029-01, and project ACM, ANR-22-CE06-0007-02. R.V. acknowledges support of ERC synergy grant SHAPINCELLFATE.\\

\textbf{Author Contributions:} MD, TVM, NL, OB, RV, TG contributed to analytical calculations. MD, TVM, TG performed numerical calculations and simulations. OB, RV, HK and TG conceived research. TVM analyzed the experiments. BG and KX  performed experiments. MD, OB, RV, HK and TG wrote and edited the manuscript. \\

\textbf{Competing Interests:} The authors declare they have no competing interests.\\

\textbf{Data and Materials Availability:} All data needed to evaluate the conclusions in the paper are present in the paper and/or the Supplementary Materials.\\

\textbf{Supplementary materials}:\\
Supplementary Text\\
Figs. S1 to S6\\
Tables S1 to S2

%%%%%%%%%%%%%%%  
\newpage 

\begin{center}
\textbf{\Large{Supplementary Materials}}
\end{center}

In the supplementary materials, we provide
\begin{itemize}
\item a detailed derivation of the formalism to calculate the splitting probabilities in $d=1$ (Section \ref{FormalismSection}),
\item a control of the approximations   (Section \ref{ApproxControls}),
\item an asymptotic analysis of the theory for scale invariant processes (fBM) (Section \ref{AsymptTheory}),
\item details on experimental methods (Section \ref{ExpDetailsSection}),
\item the proof that the theory is exact at first order  for weakly non-Markovian processes and the solution of the perturbation theory (Section \ref{Perturbation}),
\item the extension of the formalism to higher dimensions (Section \ref{2DTheory}). 
\end{itemize}

 \setcounter{table}{0}
\renewcommand{\thetable}{S\arabic{table}}

 \setcounter{figure}{0}
\renewcommand{\thefigure}{S\arabic{figure}}
 
 \setcounter{equation}{0}
\renewcommand{\theequation}{S\arabic{equation}}  

\section{Formalism for the two target problem in one dimension}
\label{FormalismSection}

\subsection{Derivation of Eq.~(1)}
In this section, we show that 
\begin{align}
x_0=\lim_{t\to\infty}[ \pi_1 \mu_1(t) +\pi_2 \mu_2(t)],\label{Eq1}
\end{align}
which is Eq.~(1) in the main text. We recall that the stochastic process is Gaussian, unbiased, with MSD $\psi(t)$, starting at $x_0$, with stationary increments, so that its covariance function is (see \textit{e.g.} (43))%Ref.~\cite{guerin16})
\begin{align}
\text{cov}(x(t),x(t'))\equiv \sigma(t,t')=\frac{1}{2}\left[-\psi(\vert t-t'\vert)+\psi(t)+\psi(t')\right]\label{StatSigma}.
\end{align}
Let us first write the exact relation
\begin{align}
x_0 = \int_0^tdt' F(t') \mathbb{E}(x(t)\vert\text{FPT}=t') +\mathbb{E}(x(t)\vert\text{FPT}>t) S(t)
 \label{9431},
\end{align}
where $F(t)$ is the density of first passage times (to reach either target 1 or target 2), $S(t)=\int_t^\infty d\tau F(\tau)$ is the survival probability, $\mathbb{E}(x(t)\vert\text{FPT}=t')$ is the average of $x(t)$ given that the first passage time (FPT) is equal to $t'$, $\mathbb{E}(x(t)\vert\text{FPT}>t)$ is the average of $x(t)$ given that the first passage is larger than $t$. We introduce the following ``trick'', for any $A>0$ and any function $g(t,t')$:
\begin{align}
\int_0^A dt \int_0^t dt' \ g(t,t')=\int_0^A dt' \int_{t'}^A dt \ g(t,t')= \int_0^A dt' \int_0^{A-t'} du \ g(t'+u,t')=\int_0^A du \int_0^{A-u} dt' \ g(t'+u,t').\label{Trick}
\end{align}
Using this property, if we integrate (\ref{9431}) over $t$ between $0$ and $A$, we obtain
\begin{align}
\int_0^A dt \ x_0 = \int_0^A du \int_0^{A-u} dt' \ F(t') \mathbb{E}(x(t'+u)\vert\text{FPT}=t')+\int_0^A dt \ \mathbb{E}(x(t)\vert\text{FPT}>t) S(t). \label{0421}
\end{align}
We consider the average trajectory $\mu(t)$ in the future of the FPT:
\begin{align}
\mu(t)=\langle x(t+\text{FPT})\rangle=\int_0^\infty dt' F(t')  \mathbb{E}(x(t'+t)\vert\text{FPT}=t').\label{DefMu}
\end{align}
Using (\ref{0421}) and (\ref{DefMu}) we obtain
\begin{align}
\int_0^A dt\  [\mu(t)-x_0]= \int_0^A du \int_{A-u}^\infty dt' \  F(t') \mathbb{E}(x(t'+u)\vert\text{FPT}=t') - \int_0^A dt \ \mathbb{E}(x(t)\vert\text{FPT}>t) S(t) \equiv Q(A),\label{DefQ}
\end{align}
where $Q(A)$ is defined by the above equation. We now wish to show that $Q(A)/A$ vanishes for large $A$. 
First, we set $v=A-u$:
\begin{align}
Q(A)=\int_0^A dv \int_{v}^\infty dt' \  F(t') \mathbb{E}(x(t'+A-v)\vert\text{FPT}=t')- \int_0^A dt \ \mathbb{E}(x(t)\vert\text{FPT}>t) S(t),
\end{align}
\textcolor{black}{Next, when $A$ goes to infinity at fixed $t',v$, the distribution of $x(t'+A \vert \text{FPT}=t')$  extends at most over a typical length $\sqrt{\kappa} A^{H}$ (found by dimensional analysis). Hence, the average distance $\mathbb{E}(x(t'+A)\vert\text{FPT}=t')$ travelled during the time $A$ cannot be very large compared to this typical length $\sqrt{\kappa} A^{H}$, whatever the value of the first passage time $t'$. We thus argue that there exists a constant $K_1$ such that, for $A\to\infty$ \textcolor{black}{(at fixed $t',v$)}, with $A\gg t'$,
\begin{align}
\vert\mathbb{E}(x(t'+A-v)\vert\text{FPT}=t')\vert <K_1 (A-v)^H<K_1A^H. 
\end{align}
%Note that the above inequality is expected to hold at long $A$. 
Hence, in the limit $A\to\infty$, we obtain
\begin{align}
\vert \int_0^A dv \int_{v}^\infty dt' \  F(t') \mathbb{E}(x(t'+A-v)\vert\text{FPT}=t')\vert <  \int_0^A dv \int_{v}^\infty dt' \  F(t') K_1 A^H.
\end{align}
Note that, although in the above integral the upper bound for $t'$ is formally equal to infinity, in practice only finite values of $t'$ matter since $F(t')$ decays very fast at times larger than $(L^2/\kappa)^{1/(2H)}$ (beyond which the probability of having missed the targets becomes exponentially small). \textcolor{black}{The same remark holds for the variable $v$. Hence, the limit $A\to\infty$ of $Q$ can be evaluated by using $A\gg t',v$ }}
Furthermore, the absolute value of $x(t)$, given that no boundaries have been reached before $t$, is necessarily less that $L$, so that $\vert\mathbb{E}(x(t)\vert\text{FPT}>t)\vert < L$. Using these arguments, we find
\begin{align}
\vert Q(A)\vert < \int_0^A dv S(v) (K_1 A^H+L) < (K_1    A^H + L) \int_0^\infty dv S(v). 
\end{align}
Note that $\langle T\rangle=\int_0^\infty dv S(v)$ is finite in our case, because for times larger than $(L^2/\kappa)^{1/(2H)}$ the random walker is almost sure to have reached one of the two targets. The above expression tells us that $Q(A)$ is at most of order $A^H$ for large $A$. Comparing with (\ref{DefQ}), this means that, for large $t$, $\mu(t)-x_0$  is at most of order   $1/t^{1-H}$. Hence, for $H<1$, $\mu(t)-x_0$ vanishes at large times. Since $\mu=\pi_1\mu_1+\pi_2\mu_2$, we obtain the result (\ref{Eq1}), which is Eq.~(1) of the main text.  

\subsection{Self-consistent equations for $\mu_1(t)$ and $\mu_2(t)$ [Derivation of Eq.~(2)]}
\label{DerivSelcCOns}
Let us consider the equation 
\begin{align}
p(0,t; y, t+\tau) = \int_0^t dt' F(t') p(0,t; y,t+\tau\vert \text{FPT}=t'),
\end{align}
which is exact for continuous non-smooth processes. $p(0,t; y, t+\tau)$ is the joint probability density of observing $x(t)=0$ and $x(t+\tau)=y$ (in the absence of any target), with $\tau>0$ and $t>0$. Next, $p(0,t; y,t+\tau\vert \text{FPT}=t')$ is the probability density of observing $x(t)=0$ and $x(t+\tau)=y$ given that the FPT (to reach any of the two targets) is $t'$, if the random walker is allowed to continue its motion after the first passage. Using Eq.~(\ref{Trick}), we obtain
\begin{align}
\int_0^A dt \ p(0,t; y, t+\tau) = \int_0^A du \int_0^{A-u} dt' F(t') p(0,t'+u; y,t'+u+\tau \vert \text{FPT}=t') \label{9548321}.
\end{align}
Now, we introduce the joint probability to observe the position $x$ at time $t$ after the FPT and $y$ at time $t+\tau$ after the FPT:
\begin{align}
p_\pi(x,t;y,t+\tau)=\int_0^\infty dt' \ F(t') \ p(x,t'+u; y,t'+u+\tau \vert \text{FPT}=t').
\end{align}
Using the trick $\int_0^{A-u}(...)=\int_0^\infty(...) - \int_{A-u}^\infty(...)$ and the above definition, Eq.~(\ref{9548321}) becomes 
\begin{align}
\int_0^A dt \ [p_\pi(0,t;y,t+\tau)-p(0,t; y, t+\tau)] = \int_0^A du \int_{A-u}^{\infty} dt' F(t') p(0,t'+u; y,t'+u+\tau \vert \text{FPT}=t'). 
\end{align}
We multiply the above equation by $y$, write $p_\pi(0,t;y,t+\tau)=\sum_{j=1}^2\pi_j q_j(0,t;y,t+\tau)$, and integrate over $y$ to obtain
\begin{align}
\int_0^A dt \ &\left[\sum_{j=1}^2 \pi_j q_j(0,t) \mathbb{E}_{\pi_j}(x(t+\tau)\vert x(t)=0)-p(0,t)\mathbb{E}(x(t+\tau) \vert x(t)=0) \right] =\nonumber\\
& \int_0^A du \int_{A-u}^{\infty} dt' F(t') p(0,t'+u \vert \text{FPT}=t')\mathbb{E}(x(t'+u+\tau)\vert x(t'+u)=0;  \text{FPT}=t')\equiv R(A),\label{DefRA}
\end{align}
where $R(A)$ is defined by the above equation, $\mathbb{E}_{\pi_j}(x(t+\tau)\vert x(t)=0)$ is the conditional average of $x(t+\tau+\text{FPT})$ given that target $j$ is reached first, and that $x(t+\text{FPT})=0$. We wish now to show that $R(A)\to0$ for $A\to\infty$. First, setting $u=A-v$ leads to:
\begin{align}
R(A) = \int_0^A dv \int_{v}^{\infty} dt' F(t') p(0,t'+A-v \vert \text{FPT}=t')\mathbb{E}(x(t'+A-v+\tau)\vert x(t'+A-v)=0;  \text{FPT}=t').  
\end{align}
For large times, we argue that $p_\pi(0,t)\sim K_0/t^H$ for some $K_0>0$, since the  distribution of positions extends over a length $t^{H}$. Hence,
\begin{align}
& p(0,t'+A-v \vert \text{FPT}=t')\underset{A\to\infty}{\sim}  \frac{K_0}{A^H}. 
\end{align}
Next, we argue that there exists a function $h(\tau)$ so that
\begin{align}
& \vert \mathbb{E}(x(\tau+t'+A-v)\vert x(t'+A-v)=0 ; \text{FPT}=t')\vert < h(\tau),
\end{align}
this is again related to the argument that the particle cannot travel an infinite distance during the time interval $\tau$, whatever the conditioning on the past is. With these arguments, we obtain
\begin{align}
\vert R(A)\vert<  K_0  \int_0^A dv \int_{v}^{\infty} dt' F(t')  \frac{1}{A^H} h(\tau)  \simeq  \frac{1}{A^H} \int_0^\infty dv S(v) h(\tau)   \hspace{1cm}(A\to\infty).
\end{align} 
We conclude that $R(A)$ vanishes  for large $A$. Therefore, taking the limit $A\to\infty$ in Eq.~(\ref{DefRA}) leads to
\begin{align}
\mathcal{H}_1(\tau)\equiv\int_0^\infty dt \ \Bigg\{& \pi_1 q_1(0,t) \left[\mu_1(t+\tau)-\mu_1(t)\frac{\sigma(t+\tau,t)}{\sigma(t,t)}\right]+\pi_2 q_2(0,t) \left[\mu_2(t+\tau)-\mu_2(t)\frac{\sigma(t+\tau,t)}{\sigma(t,t)}\right] \nonumber\\
&-p(0,t)\left[x_0-x_0\frac{\sigma(t+\tau,t)}{\sigma(t,t)}\right] \Bigg\}=0,  \label{SelfConsEq}
\end{align}
where we have assumed that the  stochastic process after hitting a target is Gaussian with the stationary covariance approximation, and we have used formulas for conditional averages of Gaussian variables, see \textit{e.g.} (62):
\begin{align}
\mathbb{E}(X\vert Y=y)= \mathbb{E}(X)-[\mathbb{E}(Y)-y]\frac{\text{cov}(X,Y)}{\text{var}(Y)}. 
\end{align}
Note that, since the process $x(t)$ has stationary increments,  the covariance $\sigma(t,t')=\text{cov}(x(t),x(t'))$ is given by Eq.~(\ref{StatSigma}). Finally, with the same reasoning one obtains the equation related to the second target:
\begin{align}
\mathcal{H}_2(\tau)\equiv \int_0^\infty dt \ \Bigg\{& \pi_1 q_1(L,t) \left\{\mu_1(t+\tau)-[\mu_1(t)-L]\frac{\sigma(t+\tau,t)}{\sigma(t,t)}\right\}+\pi_2 q_2(L,t) \left\{\mu_2(t+\tau)-[\mu_2(t)-L]\frac{\sigma(t+\tau,t)}{\sigma(t,t)}\right\} \nonumber\\
&-p(L,t)\left[x_0-(x_0-L)\frac{\sigma(t+\tau,t)}{\sigma(t,t)}\right]\Bigg\}=0.  \label{SelfConsEq2}
\end{align}
Eqs.~(\ref{SelfConsEq}) and (\ref{SelfConsEq2}) are equivalent to Eq.~(2) of the main text.  

\section{Validity control of the approximations of the theory}
\label{ApproxControls}

\begin{figure}
    \centering
    \includegraphics[width=17cm]{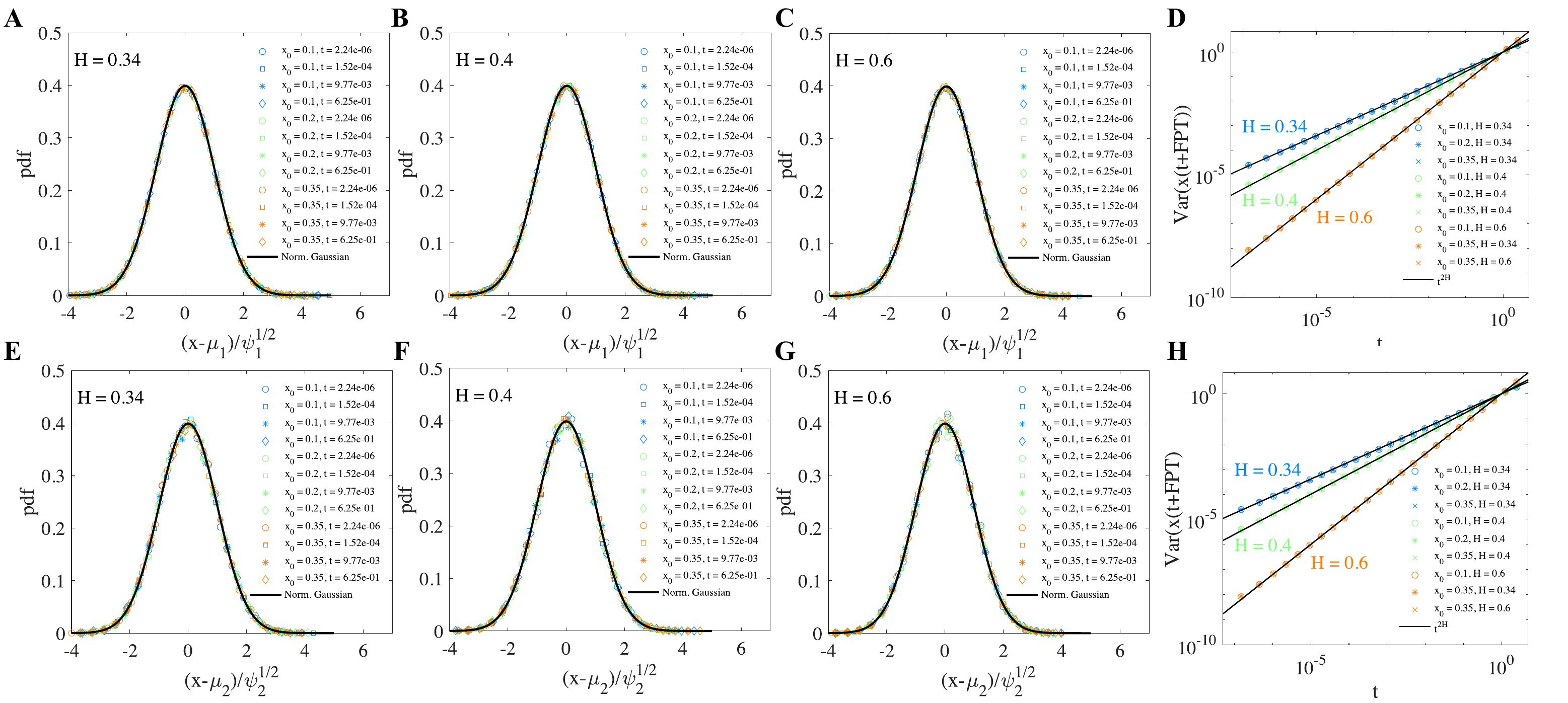}
    \caption{\textbf{Numerical check of the hypotheses of the theory.} \textbf{A, B, C}: normalized histograms of $x(t+\text{FPT})$ given that target $1$ is reached first. The stochastic process has mean square displacement $\psi(t)=t^{2H}$, with \textbf{A} $H=0.34$, \textbf{B} $H=0.4$, \textbf{C} $H=0.6$ and the targets are at $x_1=0$ and $x_2=L=1$. The time step is $\Delta t=1.5\times10^{-7}$. Symbols are the results for various values of $t$ and $x_0$ indicated in legend. The black line is a normalized Gaussian. \textbf{D}: variance   $\psi_1(t)=\text{var}[x(t+\text{FPT})]$ given that target $1$ is reached first. Symbols are simulation results for various $x_0$, the black line represents $\psi_1(t)=t^{2H}=\psi(t)$. \textbf{E, F, G, H}: same figures when target $2$ is reached first.  Note that for the lowest values of $x_0$, the amount of recorded events is less than for target 1, explaining the higher dispersion of the data. 
   }
    \label{fig:checkHist}
\end{figure}

\begin{figure}[ht!]
    \centering
    \includegraphics[width=16cm]{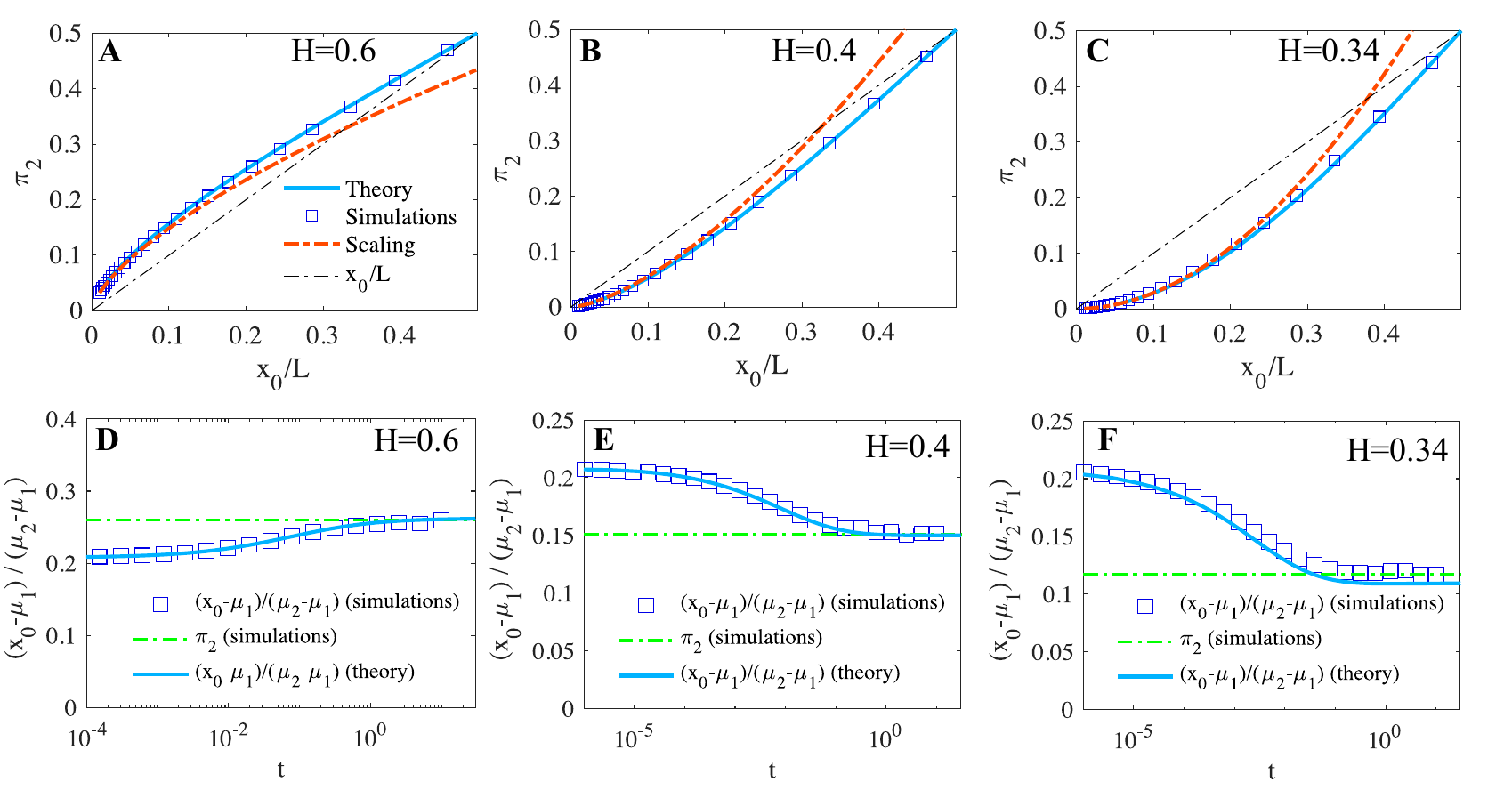}
    \caption{ \textcolor{black}{\textbf{Splitting probabilities and check of the convergence of $(x_0-\mu_1)/(\mu_2-\mu_1)$ to $\pi_2$ at large times  for the fractional Brownian motion. }Upper subfigures show the curves $\pi_2$ as a function of $x_0$, for \textbf{A}: $H=0.6$, \textbf{B}: $H=0.4$, \textbf{C}: $H=0.34$. Symbols: simulations, blue line: theory, red dashed line: scaling  (\ref{ValueAH}) with $A_H$ as given in table   \ref{TableAH}. Black dot-dashed line: pseudo-Markovian estimate obtained with $\mu_i=x_i$. \textbf{A-C}  are the same figures as in Fig.~2\textbf{A-C}, but in linear scales. Lower subfigures show  $(x_0-\mu_1)/(\mu_2-\mu_1)$ as a function of time, for  \textbf{D}: $H=0.6$, \textbf{E}: $H=0.4$, \textbf{F}: $H=0.34$. Symbols: simulations, blue line: theory, green dot-dashed line: value of $\pi_2$ in simulations. Here, $x_0/L=0.208$, and $t$ is in units of $(L^2/\kappa)^{1/(2H)}$. }}
    \label{figFBMLinear}
\end{figure}

The theory presented in Appendix \ref{FormalismSection} relies on two assumptions: (1) the process in the future of the first passage to a target $i\in \{1,2\}$ can be described as a Gaussian process, and (2) the covariance of the future of the first passage is approximated by the covariance of the original process. The validity of these assumptions is checked on Fig.~\ref{fig:checkHist}. As a direct test of the theory, we also show the splitting probabilities for the fractional Brownian motion in linear scales, see Fig.~\ref{figFBMLinear}\textcolor{black}{\textbf{A-C}. Finally, on Fig.~\ref{figFBMLinear} \textbf{D-E}, as a direct test of Eq.~(1), we show the quantity $(x_0-\mu_1)/(\mu_2-\mu_1)$, enabling one to check that it converges to $\pi_2$ for large times.}

%\vspace{2cm}

\newpage
\section{Asymptotic behavior of $\pi_2$ for scale invariant processes [Derivation of Eq.~(4)].}
\label{AsymptTheory}
Here, we consider the case $\psi(t)=\kappa\  t^{2H}$ at all times (so that, in the absence of target, $x(t)$ is a fBM). Without loss of generality, we chose the units of length and time  so that $L=\kappa=1$. We look for the value of $\pi_2$ for small $x_0$. We start with the following ansatz for the structure of the solution: 
\begin{align}
&\mu_1(t,x_0)\simeq 
\begin{cases} 
x_0\ \mu_\infty(t/x_0^{1/H}) & t\ll 1, \\
x_0 - x_0^{1/H-1}\ \chi (t) & t\gg x_0^{1/H}
\end{cases}, 
&\mu_2(t,x_0)\simeq m_2(t) \label{Ansatz},
\end{align}
where $\mu_\infty$, $\chi$ and $m_2$ are scaling functions. This ansatz is justified by the fact that, for small $x_0$ it is clear that the time to reach the closest boundary must play a role, so that $\mu_1$ must vary at the scale $x_0^{1/H}$. It must also vary at the second relevant time scale of the problem, i.e. the time to travel a distance $L$, which in our case is $1$. As shown below, $\mu_\infty$ will be the reactive trajectory for the single target problem. Furthermore,  the $x_0^{1/H-1}$ multiplicative factor of $\chi$ is imposed by the fact that $\mu_\infty(t)\simeq 1-B_\infty/t^{1-2H}$, so that the small time and large time solutions coincide when
$\chi(t)\underset{t\to0}{\simeq} B_\infty/t^{1-2H}$.
Finally, it is natural to assume that $\mu_2$ does not vary at the scale $x_0^{1/H}$. The ansatz (\ref{Ansatz}) will be justified by the fact that one can identify the equations for $\chi$, $m_2$ and $\mu_\infty$. The equation  (\ref{Eq1}) for $\pi_1$ and $\pi_2$ leads to
\begin{align}
&1-\pi_1=\pi_2=A_H \ x_0^{1/H-1}, & A_H=\lim_{t\to\infty} \frac{\chi(t)}{m_2(t)}, \label{ValueAH}
\end{align}
and thus the prefactor  $A_H$ of the scaling law for the splitting probability can be estimated from the values of $m_2$ and $\chi$ at infinity. 

The equation for $\mu_\infty$ is obtained by estimating  $\mathcal{H}_1(\tau)$ [defined in  Eq.~\eqref{SelfConsEq}] for $\tau=x_0^{1/H}\overline{\tau}$, at fixed $\overline{\tau}$ in the limit of small $x_0$, which leads to
\begin{align}
\int_0^\infty \frac{dt}{t^H} \Bigg\{ e^{-\mu_\infty^2(t)/2t^{2H}} \left[\mu_\infty(t+\overline{\tau})-\mu_\infty(t)\frac{t^{2H}+(t+\overline{\tau})^{2H}-\overline{\tau}^{2H}}{2t^{2H}}\right]-e^{-1/2t^{2H}} \left[1-\frac{t^{2H}+(t+\overline{\tau})^{2H}-\overline{\tau}^{2H}}{2t^{2H}}\right]\Bigg\}=0,
\end{align}
which is the equation for the single target problem (43) % \cite{guerin16} 
(as expected). Let us identify the equation for $\chi$. With the above scaling ansatz (\ref{Ansatz}), we estimate that $\mathcal{H}_1(\tau)$ reads in the small $x_0$ limit (at fixed $\tau$):
\begin{align}
\mathcal{H}_1(\tau)\simeq - x_0^{1/H-1} \int_0^\infty \frac{dt}{t^H}  \Bigg\{ & \left[\chi(t+\tau)-\chi(t){M_H(t,\tau)}\right]-A_H e^{-m_2^2(t)/(2t^{2H})}\left[m_2(t+\tau)-m_2(t){M_H(t,\tau)}\right]
\Bigg\}=0\label{H1pref},
\end{align}
{where
\begin{align}
M_H(t,\tau)=\frac{(t+\tau)^{2H}+t^{2H}-\tau^{2H}}{2t^{2H}}.
\end{align}
 Eq.~(\ref{H1pref}) } provides an equation for $\chi$ and $m_2$. 
Next, we note that
\begin{align}
&e^{-\frac{[1-\mu_1(t)]^2}{2\psi}}\simeq e^{-\frac{\left[1-x_0+x_0^{1/H-1}\chi(t)\right]^2}{2t^{2H}}}\simeq e^{-1/(2t^{2H})}\left(1+\frac{x_0}{t^{2H}}-\chi(t)\frac{x_0^{1/H-1}}{t^{2H}}+...\right), & (x_0\to0).
\end{align}
Hence, collecting the terms of order $x_0^{1/H-1}$ in $\mathcal{H}_2(\tau)$ [defined in Eq.~\eqref{SelfConsEq2}], we obtain
\begin{align}
\mathcal{H}_2(\tau)\simeq x_0^{1/H-1} \int_0^\infty \frac{dt}{t^H}  \Bigg\{ &
\left[-\chi(t+\tau)+\chi(t){M_H}(t,\tau)\right] e^{-1/(2t^{2H})} 
-\left(A_H+\chi(t)\frac{x_0^{1/H-1}}{t^{2H}} \right)[-1+{M_H}(t,\tau)]e^{-1/(2t^{2H})}\nonumber\\
&+A_H e^{-[1-m_2(t)]^2/(2t^{2H})}\left[ m_2(t+\tau)-1-[m_2(t)-1]{M_H}(t,\tau)\right]
\Bigg\}=0.\label{H2pref}
\end{align}
{Eqs.~(\ref{H1pref}) and (\ref{H2pref}) form } a system of two equations for $\chi$ and $m_2$ which we can solve, with the advantage that there are no parameters left (apart from $H$). We also note that we have to look for solutions with $B_\infty$ as  input, obtained by solving the equation for $\mu_\infty$, as in (43)%Ref.~\cite{guerin16}
.  Some results for $A_H$   obtained by numerically solving Eqs.~(\ref{ValueAH}), (\ref{H1pref}) and (\ref{H2pref}) are reported in table   \ref{TableAH}. 

\begin{table}
\begin{center}
\begin{tabular}{|c|c| c   |}
\hline
   $H$ & $B_\infty$ & $A_H$   \\
   \hline
   0.4 & 0.75 \ &   1.75   \\
      0.34 & 0.49 & 2.5  \\
            0.6 & 1.05  & 0.69   \\
      \hline  
\end{tabular}
\end{center}
  \caption{ Values of $B_\infty$ (calculated in (43)) and $A_H$, obtained numerically by solving \eqref{H1pref} and \eqref{H2pref}. } 
  \label{TableAH}
\end{table}

\section{Details on experimental analysis}
\label{ExpDetailsSection}
  
\begin{figure}[ht!]
    \centering
    \includegraphics[width=14cm]{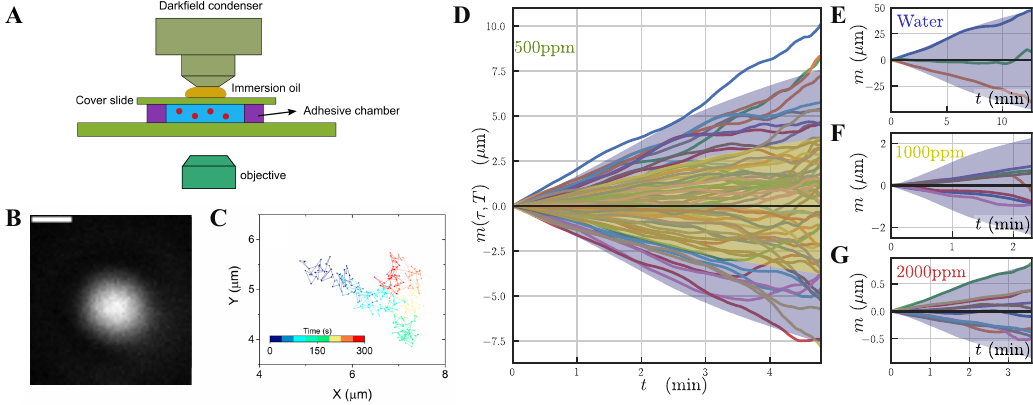}
    \caption{\textbf{Tracking of particles in viscoelastic fluids}. \textbf{A} Schematic of darkfield experiment setup. The darkfield condenser is immersed in the optical oil. The thickness of the sample is about 30 $\mu$m sealed inside a slide incubation chamber (from Bio-Rad). \textbf{B} Example of image of a particle diffusing in the polymer solution, captured using dark field microscopy. \textbf{C} Example of trajectory of a particle diffusing in 2000 ppm polymer solution. The movie was recorded at 250 frame per second (fps) using an objective magnification 100$\times$. Movie duration is about 5min. For clarity, a 1fps trajectory is plotted on the figure. \textbf{D, E, F, G} Check of the no-drift hypothesis: we represent the time averaged drift $m(t,\tau)$ defined in Eq.~(\ref{Def_m}). The purple conical regions are $\pm 2\sqrt{\text{var}(m(t,\tau))}$, which should contain $95\%$ of the observations $m(t,\tau)$ in the absence of any drift. In \textbf{D}, the yellow conical region is $\sqrt{\text{var}(m(t,\tau))}$ }
    \label{fig:trajectory}
\end{figure}

\subsubsection*{Estimation of the MSD and check of the no-drift hypothesis}
To estimate the MSD $\psi(t)$ of the tracer particle, we used the following estimator, known as the time-averaged MSD, and defined as
\begin{equation}
	\delta^2(\tau,T)=\frac{1}{T-\tau} \int_0^{T-\tau} dt \  [x (t+\tau)-x(t)]^2,
\end{equation}
where one assumes that a trajectory is observed during a time $T$. 
Obviously, $\langle\delta^2(\tau,T)\rangle=\psi(\tau)$, and the variance of $\delta^2$ can be calculated by assuming that $x(t)$ is a Gaussian process (which is suggested by the experimentally observed histograms) with stationary increments. In this case,  
\begin{align}
	\text{Var}(\delta^2  (\tau,T)) &=\int_0^{T-\tau} dz \frac{T-\tau-z}{(T-\tau)^2} \Big[ \psi(z+\tau) + \psi(|z-\tau|) - 2 \psi(z) \Big]^2.
\end{align}
This formula enables us to estimate the precision on the measurement of $\psi$ to $\pm 2\sqrt{	\text{Var}(\delta^2)/N_{\text{traj}}}$, with $N_{\text{traj}}$ the number of independent observed trajectories. We then fit the functional form of the MSD using the equations (9) and (10) in the main text. The fitting parameters are indicated in table \ref{TableMSD}.

\begin{table}
\begin{center}
{\begin{tabular}{| c | c | c | c |}
\hline
$c$ (ppm) &   $\alpha/2$ & $\gamma_0/k_BT $ (s/$\mu$m$^2$) & $\tau_0$ (s)  \\
   \hline
      0 (water) &   -  &   2.70$\pm$0.15  & -\\
   500 &   0.375$\pm$ 0.005  &   40$\pm$2  & 1.5$\pm$0.2 \\
      1000 &  0.275$\pm$0.01 &  218$\pm$12 & 2.9$\pm$0.5  \\
            2000 & 0.175$\pm$0.01  & (2.42$\pm$0.15)$\times 10^3$ & 7.0$\pm$ 1   \\
      \hline  
\end{tabular}}
\end{center}
  \caption{Parameters of the MSD for the different polymer solutions, {obtained by fitting the experimental MSDs with the functional form of Eqs.~(9) and (10) in the main text. Uncertainties indicate values for which the fit becomes unsatisfactory, given the statistical uncertainties. Since no subdiffusive regime could be observed in the experiments for water, no attempt was made to measure  values for $\tau_0$ and $\alpha$ in that case. }}    
    \label{TableMSD}
\end{table}

To determine if the deviations of $x(t)$ with respect to $x_0$ in each trajectory are related to a background drift flow, we estimated the time-averaged increments:
\begin{equation}
	m(\tau,T)=\frac{1}{T-\tau}\int_0^{T-\tau} dt   \ [x(t+\tau)-x(t)]. \label{Def_m}
\end{equation}
Under the hypothesis that $x$ satisfies the GLE equation with no drift, $\langle m(\tau,T)\rangle=0$ and its variance is 
\begin{align}
\langle m^2(t,\tau)\rangle=\int_0^{T-\tau} dz \frac{T-\tau-z}{(T-\tau)^2} \Big[ \psi(z+\tau) + \psi(|z-\tau|) - 2 \psi(z) \Big].
\end{align}
In all our experiments, about $95\%$ of the observed  values of $m(\tau)$ remained in the range $\pm 2 \langle m^2\rangle^{1/2}$ [calculated with the above formula, see Fig.~\ref{fig:trajectory}(d)-(f)], suggesting that there is no need to assume the existence of a drift to explain the data. 

\subsubsection*{Choice of parameters}

The choice of the parameters of the experiments is made to ensure that two conditions are satisfied. First, the displacement $\Delta x$ between two frames has to be small compared to $L$, so that a first passage event between two frames is not missed in the analysis. Since $\Delta x\simeq \sqrt{\kappa_0} (\Delta t)^{\alpha/2}$ for small times, where $\kappa_0=2\tau_0^{1-\alpha}k_BT/(\gamma_0\Gamma(1+\alpha))$ is the transport coefficient at small times, this condition writes
\begin{align}
\frac{L}{\Delta x}=\frac{L}{\sqrt{\kappa_0} (\Delta t)^{\alpha/2}}\gg 1.\label{Cond1}
\end{align}
We chose $L$ to be the largest possible that remains in the range where the MSD is not linear, hence at the cross-over between the subdiffusive and the diffusive regime.  
Next, the memory of the camera limits the number of images that one can acquire during one experiment. For example, at our spatial resolution, the camera can take movies of about 64,000 images. Taking $10$ movies (which represents $640$GB of data), one can record about  $n_0\simeq 640,000$ images. If we consider that one can use initial conditions separated by $2\tau_0$ as independent initial conditions, we estimate the number of first passage events potentially observable as
\begin{align}
N_\text{events}\simeq \frac{(\Delta t) \times n_0 }{2\tau_0}\gg 1\label{Cond2}
\end{align}
The temporal resolution $\Delta t$ has to be chosen so that both conditions (\ref{Cond1}) and (\ref{Cond2}) are satisfied at the same time. For example, for a polymer solution concentrated at $1000$ ppm, with $L=0.5\mu m$, with a frame rate of $245$ frames per seconds, we obtain $L/\Delta x \simeq 7$ and $N_{\text{events}}\simeq 400$, which is the order of magnitudes of the number of events available for $640$ GB of data. In our analysis we could actually observe about $500$ first passage events without the bead leaving the frame or getting close to cell surfaces. Hence, a large amount of data was used to get enough statistics and observe with enough precision non-Markovian effects in the splitting probabilities.

\section{First order perturbation theory around Brownian motion ($d=1$)}\label{Perturbation}

Here, we  show that our theory is exact at first order for weakly non-Markovian processes. Our strategy consists in identifying an exact equation defining the distribution of paths after the first passage, and then checking that, with our Gaussian approximation and the stationary covariance hypothesis, this general equation is satisfied at first order. Next, we give the explicit solution for $\pi_2$ at this order, and compare with results of the literature. 

\subsection{Exactness of the theory  at first order}\label{Perturbation_Gauss}

Let us consider a set of times $\tau_1,...,\tau_n$ and a set of positions $\{y_1,....,y_N\}$. We may write the following equation for the probability density of observing the positions $y_i$ at times $t+\tau_i$:
\begin{align}\label{renewal0}
p(x_i, t;y_1,t+\tau_1;... ;y_N,t+\tau_N )=\int_0^t\dif\tau'[&F_1(\tau')p(x_i, t;y_1,t+\tau_1;... ;y_N,t+\tau_N | 1,\tau')\nonumber\\
&+F_2(\tau')p(x_i, t;y_1,t+\tau_1;... ;y_N,t+\tau_N |2,\tau')],
\end{align}
where $p(x_i, t;y_1,t+\tau_1;... ;y_N,t+\tau_N | j,\tau')$ is the probability density of observing $x_i$ at $t$ and $y_k$ at all $t+\tau_k$, given that target $j$ was reached first at $\tau$. Using exactly the same arguments as in  Section \ref{DerivSelcCOns}, this equation leads to 
\begin{align}\label{EQh_ij0}
\sum_{j=1}^2\pi_j \int_0^{\infty}\dif t\,[q_j(y_1,t+\tau_1;...,y_N,t+\tau_N | x_i,t )q_j(x_i,t)-p(y_1,t+\tau_1;...,y_N,t+\tau_N | ; x_i,t )p(x_i,t)]=0.
\end{align}
Note that for $N=0$ we obtain:
\begin{align}\label{AlternativeEqPi}
\sum_{j=1}^2\pi_j \int_0^{\infty}\dif t\,[ q_j(x_i,t)-p(x_i,t)]=0.
\end{align}
Formally  Eq.~(\ref{EQh_ij0}) can be interpreted, in the continuous limit with $N\to\infty$ as 
\begin{align} 
\sum_{j=1}^2\pi_j \int_0^{\infty}\dif t\,[q_j([y(\tau)],t \vert x_i,t) q_j(x_i,t)-p([y(\tau)],t \vert x_i,t)p(x_i,t)]=0,
\end{align}
for all continuous paths  $[y_i(\tau)]$, satisfying $y_i(0)=x_i$, where $p([y_i(\tau)];t\vert x_0)$ is  the  joint probability density to follow this path after the  time $t$, i.e. the probability that $x(t+\tau)=y_i(\tau)$ for all $\tau$. Similarly, $q_j([y_i(\tau)],t\vert x_i,t)$ is the joint probability density that $x(t+\text{FPT}+\tau)=y(\tau)$, for all $\tau>0$, given that target $j$ was reached first, and that $x(t)=x_i$. Note that the condition $x(t)=x_i$ can be replaced by the condition that $y(0)=x_i$, and using (\ref{AlternativeEqPi})  one can write  
\begin{align} 
\mathcal{G}_i([y])\equiv\sum_{j=1}^2\pi_j \int_0^{\infty}\dif t\,\{ &[q_j([y(\tau)],t \vert y(0)=x_i)-p_s([y(\tau)] \vert y(0)=x_i) ]q_j(x_i,t)\nonumber\\
&-p(([y(\tau)],t \vert y(0)=x_i)-p_s([y(\tau)] \vert y(0)=x_i)]p(x_i,t)\}=0, 
\end{align}
where the functional $\mathcal{G}_i([y])$ is defined by the above equation for all paths $y(\tau)$, and $p_s([y(\tau)] \vert y(0)=x_i)=\lim_{t\to\infty}p([y(\tau)],t \vert y(0)=x_i)$ is the stationary probability to follow a the path $[y]$ given that it starts at $x_i$. The process corresponding to $p_s$ is  a Gaussian process of mean $x_i$ and covariance $\sigma(t,t')$. The equation $\mathcal{G}_i([y])=0$, together with Eq.~(\ref{AlternativeEqPi}) may thus be seen as a system of equation defining the distribution of paths after the first passage to one of the targets, and the splitting probabilities. Requiring that   $\mathcal{G}_i([y])=0$ for all paths $[y(\tau)]$ is equivalent to requiring that the following functional vanishes for all functions $[k(\tau)]$:
\begin{align}
\mathcal{F}_i([k])\equiv \int\mathcal{D}[y]e^{\mathrm{i}\int_0^{\infty}\dif \tau k(\tau)y(\tau)} \mathcal{G}_i([y]). 
\end{align}
In the case that the paths after the FPT are Gaussian distributed, with mean $\mu_j(\tau)$ and covariance $\gamma_j(\tau,\tau')$ if target $j$ is reached first, we can evaluate this functional:
\begin{align*}
&\mathcal{F}_i([k]) =\int_0^{\infty}\dif t \sum_{j=1}^{2}\pi_j\nonumber\\
&  \Bigg\{q_j(x_i,t) \Big[e^{\mathrm{i}\int_0^{\infty}\dif \tau k(\tau)A_{i}^{\pi_j}(t,\tau)-\frac{1}{2}\int_0^{\infty}\dif \tau\int_0^{\infty}\dif \tau' k(\tau) k(\tau')B_{i}^{\pi_j}(t,\tau,\tau')}
-e^{\mathrm{i}\int_0^{\infty}\dif \tau k(\tau)x_i-\frac{1}{2}\int_0^{\infty}\dif \tau\int_0^{\infty}\dif \tau' k(\tau) k(\tau')\sigma(\tau,\tau')}\Big]  \\
&- p(x_i,t)\Big[e^{\mathrm{i}\int_0^{\infty}\dif \tau k(\tau)A_{i}(t,\tau)-\frac{1}{2}\int_0^{\infty}\dif \tau\int_0^{\infty}\dif \tau' k(\tau) k(\tau')B_{i}(t,\tau,\tau')}
 -e^{\mathrm{i}\int_0^{\infty}\dif \tau k(\tau)x_i-\frac{1}{2}\int_0^{\infty}\dif \tau\int_0^{\infty}\dif \tau' k(\tau) k(\tau')\sigma(\tau,\tau')}\Big] 
\Bigg\},
 \end{align*}
with
\begin{align*}
&A_{i}^{\pi_j}(t,\tau)= \mu_j(t+\tau) -[\mu_j(t)-x_i]\frac{\gamma_j(t+\tau,t)}{\gamma_j(t,t)}, & A_{i}(t,\tau) = x_0 -[x_0-x_i]\frac{\sigma(t+\tau,t)}{\sigma(t,t)},\\
&B_{i}^{\pi_j}(t,\tau,\tau')= \gamma_j(t+\tau,t+\tau')-\frac{\gamma_j(t+\tau,t)\gamma_j(t,t+\tau')}{\gamma_j(t,t)},& B_{i}(t,\tau,\tau') =\sigma(t+\tau,t+\tau')-\frac{\sigma(t+\tau,t)\sigma(t,t+\tau')}{\sigma(t,t)},\\
&q_j(x_i,t)= \frac{1}{\sqrt{2\pi\gamma_j(t,t)}}\exp\left[-\frac{[x_i-\mu_j(t)]^2}{2\gamma_j(t,t)}\right],&  p(x_i,t)=\frac{1}{\sqrt{2\pi\sigma(t,t)}}\exp\left[-\frac{[x_i-x_0]^2}{2\sigma(t,t)}\right].
\end{align*}
In the following we consider a small deviation around the Brownian motion, by taking the mean-square displacement  $\psi(t)=Kt+\epsilon\psi_1(t)+\mathcal{O}(\epsilon^2)$. We may thus assume that $\mu_1$ is close to $x_1=0$ and $\mu_2$ is close to $x_2=L$, leading to the ansatz:
\begin{align}
&\mu_i(t)=x_i+\epsilon g_i(t)+\mathcal{O}(\epsilon^2), \\
&\gamma_i(t,t')=\sigma^{(0)}(t,t')+\epsilon\gamma_i^{(1)}(t,t')+\mathcal{O}(\epsilon^2), &\sigma(t,t')=\sigma^{(0)}(t,t')+\epsilon\sigma^{(1)}(t,t')+\mathcal{O}(\epsilon^2),\\
& \pi_2=(x_0+\epsilon p)/L+\mathcal{O}(\epsilon^2), &\pi_1=1-\pi_2=(L-x_0-\epsilon p)/L+\mathcal{O}(\epsilon^2). 
\end{align}
Moreover, $\sigma^{(0)}(t,t')$ is the covariance of the Brownian motion, 
\begin{align}
\sigma^{(0)}(t,t')=\kappa \min(t,t').
\end{align}
At order  $\varepsilon^0$, we find that $\mathcal{F}_i$ vanishes (as expected). At order one, introducing 
\begin{align}
C_i=e^{\text{i}\int_0^\infty d\tau k(\tau) x_i-\frac{1}{2}\int_0^{\infty}\dif \tau\int_0^{\infty}\dif \tau' k(\tau) k(\tau')\sigma^{(0)}(\tau,\tau')},
\end{align}
the functional $\mathcal{F}_i([k])$ can be recast as
$\mathcal{F}_i([k]) = \epsilon\, i \mathcal{F}_i^{(1)}([k])+\mathcal{O}(\epsilon^2),$ where
\begin{align*}
\mathcal{F}_i^{(1)}([k])= \int_0^{\infty}\dif \tau k(\tau)\left[\left(1-\frac{x_0}{L}\right)Q_{i1}(\tau)+\frac{x_0}{L}Q_{i2}(\tau)\right] +\int_0^{\infty}\dif \tau\int_0^{\infty}\dif \tau' k(\tau) k(\tau')\left[\left(1-\frac{x_0}{L}\right)R_{i1}(\tau,\tau')+\frac{x_0}{L}R_{i2}(\tau,\tau')\right],
\end{align*}
with
\begin{align}
Q_{ij}(\tau)=  \int_0^{\infty}\frac{\mathrm{i}\,\dif t}{\sqrt{2\pi \kappa t}}\left\{\left[g_j(t+\tau)-g_j(t)-(x_j-x_i)\frac{\gamma_j^{(1)}(t+\tau,t)-\gamma_j^{(1)}(t,t)}{\sigma^{(0)}(t,t)}\right]e^{-\frac{(x_i-x_j)^2}{2\kappa t}} -(x_0- x_i)\Delta(t,\tau)e^{-\frac{(x_0-x_i)^2}{2\kappa t}}
\right\},
\end{align}
where
\begin{align}
\Delta(t,\tau)=[\sigma^{(1)}(t,t)-\sigma^{(1)}(t+\tau,t)]/\sigma^{(0)}(t,t),
\end{align}
and the value of $R_{ij}$ is:
\begin{align}
R_{ij}(\tau,\tau')=&-\int_0^{\infty}\frac{\dif t}{\sqrt{8\pi \kappa t}}\Big\{ \left[\gamma_j^{(1)}(t+\tau,t+\tau')-\gamma_j^{(1)}(t+\tau,t)-\gamma_j^{(1)}(t,t+\tau')+\gamma_j^{(1)}(t,t)-\sigma^{(1)}(\tau,\tau')\right]e^{-\frac{(x_i-x_j)^2}{2\kappa t}}\nonumber\\
&-\left[\sigma^{(1)}(t+\tau,t+\tau')-\sigma^{(1)}(t+\tau,t)-\sigma^{(1)}(t,t+\tau')+\sigma^{(1)}(t,t)-\sigma^{(1)}(\tau,\tau')\right]e^{-\frac{(x_i-x_0)^2}{2\kappa t}}\Big\} . \label{DefRij}
\end{align}
Now, we show that we can find the functions $\gamma_j^{(1)}$ and $g_j$ so that $\mathcal{F}_i([k])$ vanishes for all $[k(\tau)]$ at order $\varepsilon$, meaning that our theory will be exact at order $\varepsilon$. First, we note that, since $x(t)$ has stationary increments, $\sigma$ satisfies the relation (\ref{StatSigma}) and therefore
\begin{align}
\sigma(t+\tau,t+\tau')-\sigma(t+\tau,t)-\sigma(t,t+\tau')+\sigma(t,t)=\sigma(\tau,\tau'),
\end{align}
and this is true at all orders of $\varepsilon$. Hence, if one choses  $\gamma_j^{(1)}(t,t')=\sigma^{(1)}(t,t')$, then one sees that all $R_{ij}=0$ in Eq.~(\ref{DefRij}). As a consequence, all terms that are quadratic in $k(\tau)$ in the definition of $\mathcal{F}_i^{(1)}$ vanish with this choice of $\gamma_j^{(1)}(t,t')$, meaning that the stationary covariance approximation is exact at first order. Now, for convenience let us write $g_1=-f$ and $g_2=g$. The terms of $\mathcal{F}_i^{(1)}$ that are linear in $k$ vanish if $f$ and $g$  satisfy the integral equations
\begin{align}
&\int_0^{\infty}\frac{ \dif t}{\sqrt{ t}}\left\{\frac{L-x_0}{L}[-f(t+\tau)+f(t)]+\frac{x_0}{L}e^{-\frac{L^2}{2 K t}}[g(t+\tau)-g(t)+L\Delta(t,\tau)]-x_0e^{-\frac{x_0^2}{2 K t}}\Delta(t,\tau)\right\}=0,\label{EqTraj1}\\
&\int_0^{\infty}\frac{ \dif t}{\sqrt{ t}}\left\{\frac{L-x_0}{L}e^{-\frac{L^2}{2 K t}}[-f(t+\tau)+f(t)-L\Delta(t,\tau)]+\frac{x_0}{L}[g(t+\tau)-g(t)]-(x_0-L)e^{-\frac{(L-x_0)^2}{2 K t}}\Delta(t,\tau)\right\}=0.\label{EqTraj2}
\end{align}
In the following we will obtain the functions $f(t)$ and $g(t)$ that satisfy Eqs.~\eqref{EqTraj1}-\eqref{EqTraj2}. Thus, $\mathcal{F}_i^{(1)}$ vanishes for all $[k]$, and 
we conclude that our hypothesis of Gaussianity of trajectories after the first passage, with the stationary covariance approximation, is exact at least at order $\mathcal{O}(\epsilon^1)$.

\subsection{Explicit solution of the theory at first order}
Taking derivative from Eqs.~\eqref{EqTraj1}-\eqref{EqTraj2} with respect to $\tau$ gives
\begin{align}
&\int_0^{\infty}\dif t\left[\frac{L-x_0}{L}K_1(t)f'(t+\tau)-\frac{x_0}{L}K_2(t)g'(t+\tau)\right]=x_0I_1(\tau),\nonumber\\
&\int_0^{\infty}\dif t\left[\frac{L-x_0}{L}K_2(t)f'(t+\tau)-\frac{x_0}{L}K_1(t)g'(t+\tau)\right]=(x_0-L)I_2(\tau),\label{set_o1}
\end{align}
with
\begin{align}
&K_1(t)=\frac{1}{\sqrt{t}}, \hspace{1cm} K_2(t)=\frac{e^{-\frac{L^2}{2 \kappa t}}}{\sqrt{t}},  \hspace{1cm} I_i(t)=\int_0^{\infty}\frac{\dif t}{\sqrt{t}}\left[e^{-\frac{L^2}{2 \kappa t}}-e^{-\frac{(x_i-x_0)^2}{2 \kappa t}}\right] \partial_\tau\Delta(t,\tau) \label{DefK}.
\end{align}
In order to solve the system \eqref{set_o1}, we consider the auxiliary problem:
\begin{align}
&\int_0^{\infty}\dif t\left[\frac{L-x_0}{L}K_1(t)f_1'(t+\tau)-\frac{x_0}{L}K_2(t)g_1'(t+\tau)\right]=x_0A_1e^{\mathrm{i}\omega\tau},\nonumber\\
&\int_0^{\infty}\dif t\left[\frac{L-x_0}{L}K_2(t)f_1'(t+\tau)-\frac{x_0}{L}K_1(t)g_1'(t+\tau)\right]=(x_0-L)A_2e^{\mathrm{i}\omega\tau},\label{set_o1_aux}
\end{align}
which admits the following obvious solution:
\begin{align}
&f_1'(t)= \frac{ e^{\mathrm{i}\omega t} \ L}{L-x_0}\frac{x_0A_1\tilde{K}_1(-\mathrm{i}\omega)+(L-x_0)A_2\tilde{K}_2(-\mathrm{i}\omega)}{\tilde{K}^2_1(-\mathrm{i}\omega)-\tilde{K}^2_2(-\mathrm{i}\omega)}, &g_1'(t)=e^{\mathrm{i}\omega t}\frac{L}{x_0}\frac{x_0A_1\tilde{K}_2(-\mathrm{i}\omega)+(L-x_0)A_2\tilde{K}_1(-\mathrm{i}\omega)}{\tilde{K}^2_1(-\mathrm{i}\omega)-\tilde{K}^2_2(-\mathrm{i}\omega)},
\end{align}
where $\tilde{h}(s)=\int_0^\infty dt h(s) e^{-st}$ represents the Laplace transform of a function $h$, and thus $\tilde{K}_i(-\mathrm{i}\omega)=\int_{0}^{\infty}\dif t e^{\mathrm{i}\omega t} K_i(t)$. The solution of \eqref{set_o1} is found by superposition;  writing the functions $I_i$ as a superposition of exponentials  as $I_j(\tau)=\int_{-\infty}^{\infty}\frac{\dif\omega}{2\pi}\hat{I}_j(\omega)e^{\mathrm{i}\omega\tau}$, we obtain
\begin{align}
&f'(t)=\frac{L}{L-x_0}\int_{-\infty}^{\infty}\frac{\dif\omega}{2\pi}\frac{x_0\hat{I}_1(\omega)\tilde{K}_1(-\mathrm{i}\omega)+(L-x_0)\hat{I}_2(\omega)\tilde{K}_2(-\mathrm{i}\omega)}{\tilde{K}^2_1(-\mathrm{i}\omega)-\tilde{K}^2_2(-\mathrm{i}\omega)}e^{\mathrm{i}\omega t}, \nonumber\\ 
&g'(t)=\frac{L}{x_0}\int_{-\infty}^{\infty}\frac{\dif\omega}{2\pi}\frac{x_0\hat{I}_1(\omega)\tilde{K}_2(-\mathrm{i}\omega)+(L-x_0)\hat{I}_2(\omega)\tilde{K}_1(-\mathrm{i}\omega)}{\tilde{K}^2_1(-\mathrm{i}\omega)-\tilde{K}^2_2(-\mathrm{i}\omega)}e^{\mathrm{i}\omega t}.
\end{align}
Integrating these equation over $t$  and using $f(0)=g(0)=0$ leads to
\begin{align}
&f(t)=\frac{x_0L}{L-x_0}\Phi_{11}(t)+L\Phi_{22}(t), 
&g(t)=L\Phi_{21}(t)+\frac{(L-x_0)L}{x_0}\Phi_{12}(t),\label{set_o1_formsol}
\end{align}
where
\begin{align}\label{Phi_ij_def}
\Phi_{ij}(t)=\int_{-\infty}^{\infty}\frac{\dif\omega}{2\pi}\frac{e^{\mathrm{i}\omega t}-1}{\mathrm{i}\omega}\frac{\tilde{K}_i(-\mathrm{i}\omega)\hat{I}_j(\omega)}{\tilde{K}^2_1(-\mathrm{i}\omega)-\tilde{K}^2_2(-\mathrm{i}\omega)} = \int_{-\infty}^{\infty}\frac{\dif\omega}{2\pi}\frac{e^{\mathrm{i}\omega t}-1}{\mathrm{i}\omega}\frac{\tilde{K}_i(-\mathrm{i}\omega) }{\tilde{K}^2_1(-\mathrm{i}\omega)-\tilde{K}^2_2(-\mathrm{i}\omega)} \int_{0}^{\infty}\dif x \ e^{-\mathrm{i}\omega x}I_j(x) ,
\end{align}
where the last equality follows from the definition of $\hat{I}_j(\omega)$ as the Fourier transform of $I_j(t)$ (which we consider as vanishing for negative $t$). 
Now,  let us define two functions $W_1(t)$ and $W_2(t)$ so that their Laplace transforms  read:
\begin{align}
\tilde{W}_i(s)= \frac{\tilde{K}_i(s)}{s[\tilde{K}_1^2(s)-\tilde{K}_2^2(s)]}. \label{DefWTilde}
\end{align}
Changing the variables $\omega\to-\omega$  in Eq.~(\ref{Phi_ij_def}) leads to
\begin{align} 
\Phi_{ij}(t)= -  \int_{0}^{\infty}\dif x  \int_{-\infty}^{\infty}\frac{\dif\omega}{2\pi}\frac{e^{-\mathrm{i}\omega t}-1}{\mathrm{i}\omega}\tilde{W}_i(\text{i}\omega) e^{\mathrm{i}\omega x}I_j(x)=  \int_{0}^{\infty}\dif x   [- W_i(x-t)\theta(x-t)+ W_i(x)\theta(x)] I_j(x), 
\end{align}
where in the second equality we have recognized the inverse Laplace transform. Finally, making shift of the variable $x$ in the first term  of the above expression  and using the definition of $I_j$, we arrive to
\begin{align}
&\Phi_{ij}(t)=\frac{1}{2 \kappa}\int_{0}^{\infty}\dif x \ W_i(x)  \int_{0}^{\infty}\frac{\dif y}{y^{3/2}} \left[e^{-\frac{L^2}{2 \kappa y}}-e^{-\frac{(x_j-x_0)^2}{2 \kappa y}}\right][\psi_1'(x+y+t)-\psi_1'(x+y)-\psi_1'(x+t)+\psi_1'(x)], \label{FinalResultPhiij}
\end{align}
This expression, combined with Eq.~(\ref{set_o1_formsol}), is an explicit solution for the average trajectories $f$ and $g$ if one specifies the value of $W_i$. 
The functions $W_i$ can be calculated as follows. First, using (\ref{DefK}) and (\ref{DefWTilde}) we obtain
\begin{align}
&\tilde{W}_1(s)=\frac{1}{\sqrt{\pi s}\left(1-e^{-\sqrt{8s L^2/\kappa}}\right)}=\frac{1}{\sqrt{\pi s} }\sum_{n=0}^\infty e^{-n \sqrt{8s L^2/\kappa}}, \\
&\tilde{W}_2(s)=\frac{e^{-\sqrt{2s L^2/\kappa}}}{\sqrt{\pi s}\left(1-e^{-\sqrt{8s L^2/\kappa}}\right)}=\frac{1}{\sqrt{\pi s} }\sum_{n=0}^\infty e^{-(n+1/2) \sqrt{8s L^2/\kappa}},
\end{align}
where we have used geometric series in order to identify the inverse Laplace transforms of $\tilde{W}_i(s)$, leading to:
\begin{align}
&W_1(t)=\sum_{n=0}^\infty\frac{e^{-2L^2n^2/(K t)}}{\pi\sqrt{t}}=\frac{ \vartheta_3\left(0,e^{-\frac{2 L^2}{K t}}\right)+1}{2 \pi\sqrt{t} }, 
&W_2(t)=\sum_{n=0}^\infty\frac{e^{-2L^2(n+1/2)^2/(K t)}}{\pi\sqrt{t}}=\frac{\vartheta _2\left(0,e^{-\frac{2 L^2}{ K t}}\right)}{2  \pi  \sqrt{t}}.\label{W2}
\end{align}
Here, $\vartheta_k(\cdot,\cdot)$ is the Jacobi theta function of the $k$th kind. 
Finally, let us determine now the splitting probability $\pi_2=x_0/L+\epsilon\pi_2^{(1)}+\mathcal{O}(\epsilon^2)$, at order $\epsilon^1$ we obtain from Eq.~(1) in the main text
\begin{align*}
\pi_2^{(1)}L=\lim_{t\to\infty} [\pi_1^{(0)}f(t)-\pi_2^{(0)}g(t)].
\end{align*}
Using the above results, we obtain
\begin{align}\label{pi2_1int}
\pi_2^{(1)}&=\int_{0}^{\infty}\hspace{-0.3cm}\int_{0}^{\infty}\hspace{-0.1cm}\frac{\dif x\dif y}{\sqrt{y^3x}} \frac{ 1+\vartheta_4\left(0,e^{-\frac{L^2}{2 \kappa x}}\right)}{4\pi \kappa } \left\{\frac{x_0}{L}\left[e^{-\frac{L^2}{2\kappa y}}-e^{-\frac{x_0^2}{2\kappa y}}\right]-\left(1-\frac{x_0}{L}\right)\left[e^{-\frac{L^2}{2 \kappa y}}-e^{-\frac{(L-x_0)^2}{2\kappa y}}\right]\right\}[\psi_1'(x+y)-\psi_1'(x)],
\end{align}
where we have used the relation $\vartheta_3(0,q)-\vartheta_2(0,q)=\vartheta_4(0,q^{1/4})$.

\subsection{Examples}

\textit{Fractional Brownian motion.} This process is characterized by $\psi(t)=\kappa t^{2H}=\kappa t+\epsilon\psi_1(t)+\mathcal{O}(\epsilon^2)$, so that for $H=1/2+\epsilon$ we have $\psi_1(t)=2\kappa t\ln t$. We use the notation $u=x_0/L$. The splitting probability has the structure   
\begin{align}\label{pi2_1terms}
\pi_2^{(1)}=Q(u)-Q(1-u), \text { with } Q(u)\equiv\frac{u}{\pi}\int_{0}^{\infty}\frac{\dif x}{\sqrt{x}}\sum_{k=0}^{\infty}(-1)^ke^{-\frac{k^2}{x}}\int_{0}^{\infty}\frac{\dif y}{y^{3/2}}\left[e^{-\frac{u^2}{y}}-e^{-\frac{1}{y}}\right]\ln\frac{x+y}{x}, 
\end{align}
where we have used  $\vartheta_4(0,e^{-\frac{1}{x}})=1+2\sum_{k=1}^{\infty}(-1)^ke^{-\frac{k^2}{x}}$.
$Q(u)$ can be calculated by integrating first over $y$, and then over $x$, and finally summing over $k$, leading to
\begin{align}
Q(u) 
&=4u(1-\ln u)+(1-2u)\left[12\ln A-\frac{7}{3}\ln 2\right]-2\ln\pi+8\left[\psi^{(-2)}\left(\frac{u+2}{2}\right)-\psi^{(-2)}\left(\frac{u+1}{2}\right)\right],\label{QFBM}
\end{align}
where $A\approx1.28243$ is the Glaisher-Kinkelin constant and  $\psi^{(-2)}(\xi)=\int_0^{\xi}\mathrm{d}z\ln\Gamma(z)$ is the generalized polygamma function. We note that the constant term can be reformulated as $\ln(A^{12}2^{-\frac{7}{3}}\pi^{-2})=8[\psi^{(-2)}(1/2)-\psi^{(-2)}(1)]$, hence $Q(0)=0$. 
Note that the result (\ref{QFBM}) were obtained in (55) based on other methods.

\textit{Bi-diffusive process.} This process is characterized by $\psi(t)=\kappa t+A(1-e^{-t/\tau})=\kappa t+\epsilon\psi_1(t)+\mathcal{O}(\epsilon^2)$, so that $\psi_1(t)= \kappa \tau(1-e^{-t/\tau})$. With this, introducing the dimensionless variables $\ell=L\sqrt{2/(K\tau)}$ and $u=x_0/L$, we obtain
\begin{align*}
f(t)&=\frac{L[\cosh((1-u)\ell)-(1-u)\cosh(\ell)-u]}{(1-u)\ell\sinh(\ell)}(1-e^{-t/\tau}), \hspace{0.5cm}
g(t)=\frac{L[\cosh(u\ell)-u\cosh(\ell)-(1-u)]}{u\ell\sinh(\ell)}(1-e^{-t/\tau}),\\
\pi_2^{(1)}&=\frac{[\cosh((1-u)\ell)-\cosh(u\ell)+(1-2u)(1-\cosh(\ell))]}{\ell\sinh(\ell)}.
\end{align*}

\section{Generalization of the theory to $d\ge1$}
\label{2DTheory}

Here we consider the case of spatial dimension $d\ge1$. {Let us first derive general formulas for the splitting probability for a random walker of position $\ve[r](t)=(x_1(t),x_2(t),...,x_d(t))$ which has non-smooth trajectories. To the difference of the $d=1$ case, we assume that the dynamics occurs in confinement, so that the pdf of positions, which we call $p^c(\ve[r],t)$, tends to a stationary value $p_s^c$ for $t\to\infty$.}
The centers of the targets are located at $\ve[r]_1$ and $\ve[r]_2$, these targets have radius $a$ and are inside a large confining volume $V$. We first write the generalized renewal equation:
\begin{align}
p^c(\ve[r]_i,t) = \int_0^t dt' F(t') p^c(\ve[r]_i,t \vert \text{FPT}=t'),
\end{align}
where $p^c(\ve[r],t\vert \Omega )$ represents the probability density of $\ve[r]$ at $t$ given the event $\Omega$. Note that $p^c$ is defined in confined space and tends at large time to the stationary value $p_s^c(\ve[r])$.  Substracting $p_s^c$ on both sides of the above equation leads to
\begin{align}
p^c(\ve[r]_i,t)-p_s^c(\ve[r]_i) = \int_0^t dt' F(t') [p^c(\ve[r]_i,t \vert \text{FPT}=t') -p_s^c(\ve[r]_i)] - \int_t^\infty d\tau p_s^c(\ve[r]_i)F(\tau) \label{R542}. 
\end{align}
Now,   the probability density $p_\pi^c$ to observe the position $\mathbf{r}$ at time $t$ after the FPT is defined as
\begin{align}
p_\pi^c(\ve[r],\tau)=\int_0^\infty dt\ p^c(\ve[r],t+\tau\vert \text{FPT}=\tau)F(\tau).
\end{align}
We note that, using Eq.~(\ref{Trick}) with $A=\infty$ and $g(t,t')=F(t') [p^c(\ve[r]_i,t \vert \text{FPT}=t')-p_s^c(\ve[r]_i)]$, we obtain 
\begin{align}
\int_0^\infty dt \int_0^t dt' F(t') [p^c(\ve[r]_i,t \vert \text{FPT}=t') -p_s^c(\ve[r]_i)] &=\int_0^\infty du \int_0^t dt' F(t') [p^c(\ve[r]_i,t'+u \vert \text{FPT}=t') -p_s^c(\ve[r]_i)]\nonumber \\
&= \int_0^\infty du \ [p_\pi^c(\ve[r]_i,u) -p_s^c(\ve[r]_i)] 
\end{align}
Noting also that $\langle T\rangle = \int_0^\infty dt \int_t^\infty F(t')dt'$, we see that integrating  (\ref{R542}) over $t\in]0,\infty [$ leads to the exact relation 
\begin{align}
\langle T\rangle p_s^c(\ve[r]_i) = \int_0^\infty dt\  [p_\pi^c(\ve[r]_i,t)-p^c(\ve[r]_i,t)].\label{95441}
\end{align}
Partitioning over first passage to each of the targets leads to
\begin{align}
p_\pi^c(\ve[r],t)=\pi_1 q_1^c(\ve[r],t)+\pi_2 q_2^c(\ve[r],t),
\end{align}
where $q_j^c(\ve[r],t)$ is the probability density function (pdf) of $\ve[r]$ at a time $t$ after the first passage to target $j$. Hence, Eq.~(\ref{95441}) leads to a system of equations for $\pi_1,\pi_2,\langle T\rangle$ which is completed by the relation $\pi_1+\pi_2=1$, so that
\begin{align}
&\pi_1=\frac{h_{22}-h_{12}}{h_{22}+h_{11}-h_{21}-h_{12}} =1-\pi_2 \label{pi_all},\\
& \frac{\langle T\rangle}{V}= \frac{ h_{11} h_{22} -h_{12} h_{21} }{ h_{22} +h_{11} -h_{21} -h_{12}},\label{mfpt}
\end{align}
where 
\begin{equation}\label{h_ij}
h_{ij}=\int_0^{\infty}d t[q_j^c(\ve[r]_i,t)-p^c(\ve[r]_i,t)],  
\end{equation}
We stress that the above relations are exact for non-smooth processes whose pdf $p^c$ reaches a steady state $p_s^c$, \textcolor{black}{as long as one uses propagators in confined space in these expressions.} 

\begin{figure}[t!]
	\includegraphics[width=16cm]{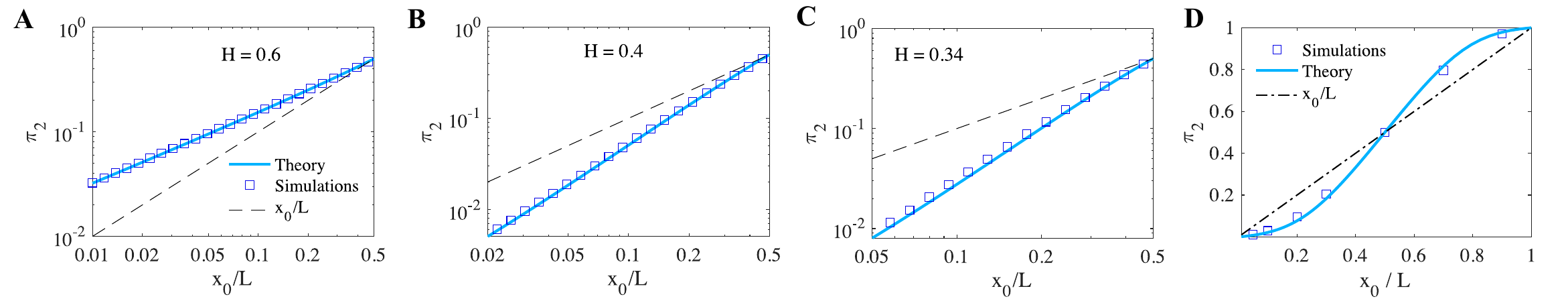}
	\caption{\textbf{Splitting probabilities for the fractional Brownian motion and the bidiffusive process in one dimension. } 
	The represented quantities are the same as in Fig.~2\textbf{A-D} when $\pi_2$ is evaluated with (\ref{pi_all}) rather than with Eq.~(2). 
	\label{FigSplittingWithhij}}
\end{figure}

To proceed further, we need to evaluate the propagators entering into the $h_{ij}$ terms. 
We use the following assumptions. First, we assume the boundaries of the volume are far enough so that all propagators can be evaluated in infinite space, the results will be valid for $V\to\infty$ when all the other parameters (distances to the targets, their radii, etc) are kept constant. 
{In this case, we can replace the propagators $p^c$ in the expressions of $h_{ij}$ by their values in infinite space, $p^c\simeq p$. We also assume that, in this large volume limit, the trajectories $x_d(t)$ satisfy the same properties as in $d=1$: they are Gaussian, continuous, non-smooth, with stationary increments. Assuming isotropic walk leads to $\text{cov}[x_i(t),x_j(t)]=\delta_{ij}\psi(t)$ and then $\text{cov}[x_i(t),x_j(t')]=\delta_{ij}\sigma(t,t')$. We again assume that $\psi(t)\sim t^{2H}$ at long times. Note that since the MSD $\psi(t)$ is defined in free space, this divergence at long times is not contradictory with the fact that the random walk occurs in confined space; in other words the MSD in confined space will saturate at times at which the boundaries can be reached. } Of note, when $d=1$, Eq.~(\ref{pi_all}) provides an alternative evaluation of $\pi_2$ to Eq.~(1); it turns out that using both equations  (\ref{pi_all}) or (1) lead to the same results, as controlled on  Fig.~\ref{FigSplittingWithhij}.

Second, we use here {a} decoupling approximation, by assuming that $q_i(\ve[r],t)$ is equal to the pdf of positions after the FPT to target $i$ when only target $i$ is present, in the single target problem. {In this approximation, one neglects the fact that some trajectories may have touched the other target first and should not be taken into account to calculate $q_i(\ve[r],t)$, we expect that this approximation is valid when the distance between the targets is not too small.}

\begin{figure}[t!]
	\includegraphics[width=5cm]{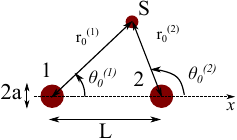}
	\caption{\textbf{Notations for the geometry of the initial position and the target locations in the 2D problem. }
	\label{FigGeom2D}}
\end{figure}

\begin{figure}[t!]
	\includegraphics[width=\linewidth]{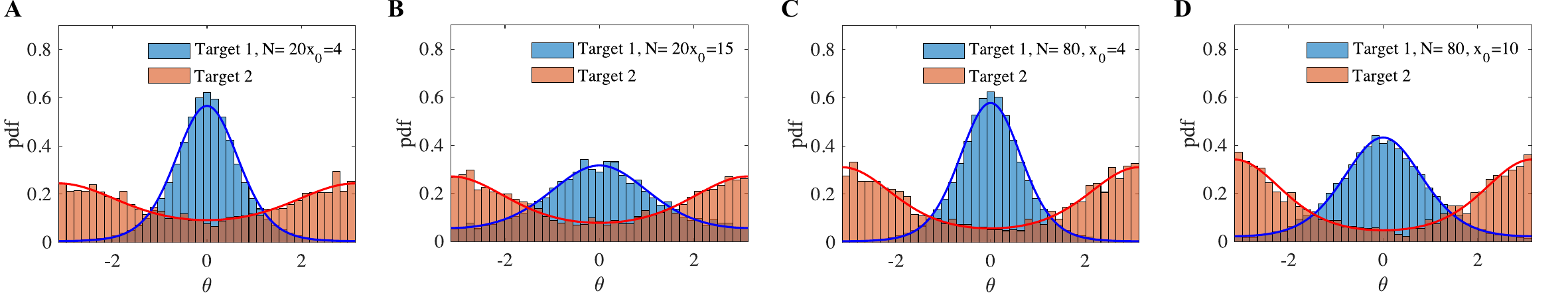}
	\caption{{\textbf{Control of the validity of the ansatz of Eq.~(\ref{ansatz}) for the system of Fig.~4(b) where the random walker is the first monomer of a Rouse polymer of $N$ monomers.} We show the histograms of the entrance angle $\theta$ at the first passage to target $1$ or $2$, and a fit with the form (\ref{ansatz}). Parameters: \textbf{A}: $N=20$, $x_0=4$; \textbf{B}: $N=20$, $x_0=15$; \textbf{C}: $N=80$, $x_0=4$ ; \textbf{D}}: $N=80$, $x_0=10$.
	\label{FigHist}}
\end{figure}

Let us now focus on the $d=2$ case, the notations describing the geometry of the targets and the initial position are specified on Fig.~\ref{FigGeom2D}. 
We focus temporarily on the single target problem, focusing on target $i$. 
The initial position is $\ve[r]_0$ and the angle between $(\ve[r]_0-\ve[r]_i)$ with respect to the $x$ axis is $\theta_0^{(i)}$. 
If the  position at the target surface when the target is hit is $\ve[r]_s$, we define $\theta_i$ the angle between $\ve[r]_s-\ve[r]_i$ and the $x$ axis. 
We call $\Pi_i(\theta_i)$ the pdf of $\theta_i$. We use the additional approximation that, after a FPT event with entrance angle $\theta_i$ (and entrance position $\ve[r]_s(\theta_i)$), the average trajectory after the FPT is $\ve[r]_i+\mu_i(t)\hat{\ve[u]}(\theta)$, where  $\hat{\ve[u]}(\theta)$    is the unit vector oriented in the direction $ \theta$. Here $\mu_i(t)$ is assumed to be independent of $\theta_i$. In these conditions, in the single target problem with only target $i$, one has 
\begin{align}
\frac{\langle T\rangle_i}{V} = \int_0^\infty dt [q_i(\ve[r]^*,t)-p(\ve[r]^*,t)] = \int_0^\infty dt  \left\{ \int_{-\pi}^\pi d\theta_i \ \Pi_i(\theta_i) \frac{e^{-\frac{[\ve[r]^*-\mu_i\hat{\ve[u]}(\theta_i)]^2}{2\psi}}}{2\pi\psi(t)}-\frac{e^{-\frac{[\ve[r]^*-x_0\hat{\ve[u]}(\theta_0^{(i)})]^2}{2\psi(t)}}}{2\pi\psi(t)}\right\} ,
\end{align}
for any $\ve[r]^*$ inside the target. Here, $\langle T\rangle_i$ is the mean FPT to target $i$ in the single target problem. Taking $\ve[r]^*=a \ e_r(\theta^*)$, multiplying by $\cos(\theta^*-\theta_0^{(i)})$  and integrating over $\theta^*$ leads 
\begin{align}
0=\int_0^\infty dt  \int_0^{2\pi} d\theta^* \cos(\theta^*-\theta_0^{(i)}) \left\{ \int_{-\pi}^\pi d\theta_i \ \Pi_i(\theta_i) \frac{e^{-\frac{a^2+\mu_i^2-2a\mu_i\cos(\theta_i-\theta^*)}{2\psi(t)}}}{2\pi\psi(t)}-\frac{e^{-\frac{a^2+x_0^2-2ax_0 \cos(\theta^*-\theta_0^{(i)})}{2\psi(t)}}}{2\pi\psi(t)}\right\} .
\end{align} 
The integration over $\theta^*$ leads to
\begin{align}
0= \int_0^\infty dt     \left\{ \int_{-\pi}^\pi d\theta_i \ \Pi_i(\theta_i)  \frac{e^{-\frac{(a-\mu_i)^2}{2\psi(t)}}}{\psi(t)}\tilde{I}_1\left(\frac{a\mu_i}{\psi} \right)\cos(\theta_i-\theta_0^{(i)})- \frac{e^{-\frac{(a-x_0)^2}{2\psi(t)}}}{\psi(t)}\tilde{I}_1\left(\frac{ax_0}{\psi} \right) \right\} ,
\end{align} 
where $\tilde{I}_1(x)=e^{-x}I_1(x)$, with $I_1$ the modified Bessel function of first kind. As a consequence, we find  
\begin{align}
\langle\cos(\theta_i-\theta_0^{(i)})\rangle_{\Pi_i}= \frac{\int_0^\infty dt  \ \frac{e^{-\frac{(a-x_0)^2}{2\psi(t)}}}{\psi(t)}\tilde{I}_1\left(\frac{ax_0}{\psi} \right) }{\int_0^\infty dt   \ \frac{e^{-\frac{(a-\mu_i)^2}{2\psi(t)}}}{\psi(t)}\tilde{I}_1\left(\frac{a\mu_i}{\psi} \right)}  \label{AvCos}.
\end{align}
Now, we will use the ansatz 
\begin{align}
\Pi_i(\theta_i)=\frac{e^{\alpha_i\cos(\theta_i-\theta_0^{(i)})}}{2\pi I_0(\alpha_i)}, \label{ansatz}
\end{align}
{which is one of the most simple function that is $2\pi-$periodic and always positive, and is also suggested by simulation results, see Fig.~\ref{FigHist} . We obtain}
\begin{align}
\langle \cos(\theta-\cos\theta_0^{(i)})\rangle_{\Pi_i} =\frac{I_1(\alpha_i)}{I_0(\alpha_i)} \label{alpha_i}.
\end{align}
As a consequence, for any position $\ve[r]$ so that the angle between $\ve[r]-\ve[r]_i$ and the $x$ axis is $\theta$, at a distance $r$ from target $i$, one has
\begin{align}
q_i(\ve[r]=\ve[r]_i+r \hat{\ve[u]}(\theta),t)=\int_0^{2\pi}d\theta_i \Pi_i(\theta_i) \frac{e^{-\frac{[r \hat{\ve[u]}(\theta) - \mu_i \hat{\ve[u]}(\theta_i)]^2}{2\psi}}}{2\pi\psi} =\int_0^{2\pi}d\theta_i   \frac{e^{\alpha_i\cos(\theta_i-\theta_0^{(i)})}}{2\pi I_0(\alpha_i)} \times \frac{e^{-\frac{ r^2+\mu_i^2-2r\mu_i\cos(\theta-\theta_i)}{2\psi}}}{2\pi\psi} \nonumber\\
= \int_0^\infty \frac{dt}{2\pi \psi }  \frac{e^{-\frac{r^2+\mu_i^2}{2\psi}}\ I_0\left[\sqrt{\left(\alpha_i+\frac{r\cos(\theta-\theta_0^{(i)})\mu_i}{\psi} \right)^2+\left(\frac{r\sin(\theta-\theta_0^{(i)})\mu_i}{\psi} \right)^2}\right]}{I_0(\alpha_i)}.
\end{align}
where we have used $\int_0^{2\pi}dt e^{a \cos t+b\sin t}=2\pi I_0(\sqrt{a^2+b^2})$. The procedure to evaluate $\pi_2$  is the following: first, we evaluate $\mu_1(t)$ and $\mu_2(t)$ for the single target problem, see (43)%Ref.~\cite{guerin16}
, then we find $\alpha_1$ and $\alpha_2$ using (\ref{alpha_i}) and  (\ref{AvCos}), and finally we use the above expression to evaluate the $h_{ij}$ in Eqs.~(\ref{pi_all}) and (\ref{h_ij}).

\end{document}